\documentclass[iop,apj,twocolappendix]{emulateapj}
\usepackage{apjfonts}
\usepackage{amssymb,epsfig,graphicx}
\usepackage{mathptmx}
\usepackage{amsmath}

\newcommand{\OII}{\hbox{[O\,{\sc ii}]}}
\newcommand{\OIII}{\hbox{[O\,{\sc iii}]}}
\newcommand{\NII}{\hbox{[N\,{\sc ii}]}}

\newcommand{\SIII}{\hbox{[S\,{\sc iii}]}}

\newcommand{\HII}{\hbox{H\,{\sc ii}}}

\shortauthors{Mao, Lin, \& Kong}
\shorttitle{\HII~Regions in NGC~2403}

\begin{document}

\title{\textbf{INTERNAL VARIATIONS IN EMPIRICAL OXYGEN ABUNDANCES FOR GIANT \HII~REGIONS \\
IN THE GALAXY NGC~2403}}

\author{Ye-Wei Mao\altaffilmark{~1}, Lin Lin\altaffilmark{~2}, Xu Kong\altaffilmark{~3,4} \vspace*{1.2mm}}

\affil{$^1$~Center for Astrophysics, GuangZhou University, GuangZhou 510006, China; \textbf{ywmao@gzhu.edu.cn}}
\affil{$^2$~Shanghai Astronomical Observatory, Chinese Academy of Sciences, ShangHai 200030, China; \textbf{linlin@shao.ac.cn}}
\affil{$^3$~CAS Key Laboratory for Research in Galaxies and Cosmology, Department of Astronomy, University of Science and Technology of China, \\HeFei 230026, China; \textbf{xkong@ustc.edu.cn}}
\affil{$^4$~School of Astronomy and Space Sciences, University of Science and Technology of China, HeFei, 230026, China \vspace*{0.8mm}}

\begin{abstract}
This paper presents a spectroscopic investigation of 11 \HII~regions in the nearby galaxy NGC~2403. The \HII~regions are observed with a long-slit spectrograph mounted on the 2.16 m telescope at XingLong station of National Astronomical Observatories of China. For each of the \HII~regions, spectra are extracted at different nebular radii along the slit-coverage. Oxygen abundances are empirically estimated from the strong-line indices $R23$, $N2O2$, $O3N2$, and $N2$ for each spectrophotometric unit, with both observation- and model-based calibrations adopted into the derivation. Radial profiles of these diversely estimated abundances are drawn for each nebula. In the results, the oxygen abundances separately estimated with the prescriptions on the basis of observations and models, albeit from the same spectral index, systematically deviate from each other; at the same time, the spectral indices $R23$ and $N2O2$ are distributed with flat profiles, whereas $N2$ and $O3N2$ exhibit apparent gradients with the nebular radius. Because our study naturally samples various ionization levels, which inherently decline at larger radii within individual \HII~regions, the radial distributions indicate not only the robustness of $R23$ and $N2O2$ against ionization variations but also the sensitivity of $N2$ and $O3N2$ to the ionization parameter. The results in this paper provide observational corroboration of the theoretical prediction about the deviation in the empirical abundance diagnostics. Our future work is planned to investigate metal-poor \HII~regions with measurable $T_\mathrm{e}$, in an attempt to recalibrate the strong-line indices and consequently disclose the cause of the discrepancies between the empirical oxygen abundances.
\end{abstract}

\keywords{\HII~regions - ISM: lines and bands - galaxies: abundances - galaxies: individual (NGC~2403) - galaxies: ISM}

\section{\textbf{INTRODUCTION}}\label{Sec_intro}

H\,{\sc ii} regions are gaseous nebulae ionized by high-energy radiation from young massive stars associated with star formation. Therefore, they are natural laboratories of star-forming activities and photoionization processes. The most striking characteristics of \HII~regions are a wealth of hydrogen and metal emission lines in spectra. These spectral lines, coded by underlying physical properties, provide powerful insight into the nature of ionized nebulae and ionizing sources.

Oxygen abundance is a physical parameter usually derived from spectral lines at optical bands. Direct diagnostics of the oxygen abundance depend on estimates of electron temperature ($T_\mathrm{e}$) by measuring auroral lines such as $\OIII\lambda4363$, $\NII\lambda5755$, and $\SIII\lambda6312$ \citep{1992AJ....103.1330G, 1993ApJ...411..655S, 2003ApJ...591..801K}. However, since the strength of the auroral lines rapidly decreases with increasing metallicity, they are always very weak and even undetectable in metal-rich objects. Therefore, the $T_\mathrm{e}$ method is usually applicable only to low metallicities ($< 0.5 Z_\odot$). Fortunately, a series of strong-line (collisionally excited) indices in spectra have been empirically calibrated as alternative tracers to the oxygen abundance, including $R23$ \citep[ $\equiv$ $\log((\OII\lambda3727+\OIII\lambda\lambda4959,5007)/\mathrm{H}\beta)$;][]{1979MNRAS.189...95P, 1991ApJ...380..140M, 1999ApJ...514..544K}, $N2$ \citep[ $\equiv$ $\log(\NII\lambda6583/\mathrm{H}\alpha)$;][]{1998AJ....116.2805V, 2002MNRAS.330...69D, 2004MNRAS.348L..59P}, $O3N2$ \citep[ $\equiv$ $\log((\OIII\lambda5007/\mathrm{H}\beta)/(\NII\lambda6583/\mathrm{H}\alpha))$;][]{1979A&A....78..200A, 1999ApJ...516...62D, 2004MNRAS.348L..59P}, and $N2O2$ \citep[ $\equiv$ $\log(\NII\lambda6583/\OII\lambda\lambda3727)$;][]{2000ApJ...542..224D, 2002ApJS..142...35K, 2007ApJ...656..186B}. The empirical diagnostics have been commonly adopted to probe metallicity for star-forming galaxies as a whole \citep[e.g.,][]{2008ApJ...681.1183K} or star-forming regions in galaxies \citep[e.g.,][]{2010ApJS..190..233M}. Nevertheless, there appears to be a considerable discrepancy between estimates from different strong-line indices, or from the same index but through different approaches to calibrations (i.e., on the basis of observations or models), which complicates the usage of the empirical diagnostics \citep[see comparisons in][and the references therein]{2008ApJ...681.1183K}. Detailed reasons for this discrepancy are still unclear, but according to photoionization models, some other parameter such as the ionization parameter in addition to the oxygen abundance is likely to have an effective impact on the strong lines \citep{2002ApJS..142...35K}; at the same time, some undefined problems in observation- or model-based calibrations are suspected to introduce nonnegligible bias in the abundance estimates \citep{2003ApJ...591..801K}. Readers are referred to \citet{2017PASP..129d3001P} for a detailed review of the direct determinations and the empirical estimations of chemical abundances in nebulae.

At present, most spectroscopic observations of \HII~regions in galaxies are taken on a spatially unresolved basis, yet these kinds of studies have not provided observational evidence of the theoretical predictions of the reasons for the discrepancies between the empirical abundances. Spatially resolved measurements, by contrast, have potential for disclosing features of the additional parameters underlying the strong-line indices. For instance, given that the degree of ionization inherently decreases from the center to the edge in an ionized nebula, dissecting individual \HII~regions is a natural approach to sampling various ionization states. Notwithstanding, due to a requirement for high spatial resolution, the spatially resolved investigations are still few in number and limited to very nearby objects \citep{2000ApJ...539..687O, 2010MNRAS.402.1635R, 2011MNRAS.413.2242M}. In this situation, the sample of this kind of study could be very small or even contain only one nebula, which restricts the obtained results to only special cases. In order to draw more universal conclusions, a larger sample of \HII~regions measured in a spatially resolved way are necessary.

The work presented in this paper is a spectrophotometric investigation of giant \HII~regions in the nearby galaxy NGC~2403. Spatially unresolved observations of \HII~regions in NGC~2403 have been taken in several studies \citep{1985ApJS...57....1M, 1997ApJ...489...63G, 1999ApJ...513..168G, 1998AJ....116.2805V, 1999ApJ...510..104B, 2013ApJ...775..128B}. \citet{2010ApJS..190..233M} has compiled most of these integrated measurements and resulted in a typical oxygen abundance of $\sim 8.80$ in $12 + \log(\mathrm{O/H})$ with a gradient of $\sim -0.024$ dex per arcmin by taking a model-based prescription into derivation, and of $\sim 8.30$ in $12 + \log(\mathrm{O/H})$ with a gradient of $\sim -0.029$ dex per arcmin through an observational diagnostic. In our work, the observations are taken with a long-slit spectrograph. Each of the \HII~regions is spatially resolved into several units on the path covered by the slit, where optical spectra are extracted. The goal of this study is to examine the deviation between the various diagnostics of the oxygen abundance, corroborate the theoretical interpretation of the abundance discrepancy, and thereby supplement previous single-nebula studies. As an ideal target for our work, NGC~2403 is a late-type spiral (Sc III) galaxy and is rich in \HII~regions \citep{1983AJ.....88..296H, 1990A&A...237...23S, 1997ApJ...489...63G}, which guarantees sufficient candidates of proper objects for sample selection. The location of NGC~2403 is of particular advantage for the long-slit spectrometry. The proximity of NGC~2403 \citep[$\sim 3.5$ Mpc,][]{2006MNRAS.369.1780V} brings a large angular size of this galaxy \citep[$D_{25} \sim 21\arcmin.9 \times 12\arcmin.3$ for NGC~2403 and $> 10\arcsec$ for many of its interior \HII~regions,][]{1999AJ....117.1249D}, which ensures enough spatial resolution for spatially resolved measurements; on the other hand, NGC~2403 is located slightly further than very nearby objects in the local group, which allows the spectrographic slit to cover not only multiple objects at one exposure but also a high quality of background at blank areas.

The remainder of this paper is outlined as follows. In Section \ref{Sec_obs}, we describe spectroscopic observations of the \HII~regions selected in NGC~2403 as the sample. In Section \ref{Sec_data}, we present data-processing procedures. The spectrophotometric data are used to derive empirical oxygen abundances and other relevant results, which are presented in Section \ref{Sec_res}. Finally, we interpret the presented results and discuss their implications in Section \ref{Sec_disc}.

\section{\textbf{OBSERVATIONS}}\label{Sec_obs}

\begin{figure*}[!ht]
\centering
\includegraphics[width=2.0\columnwidth]{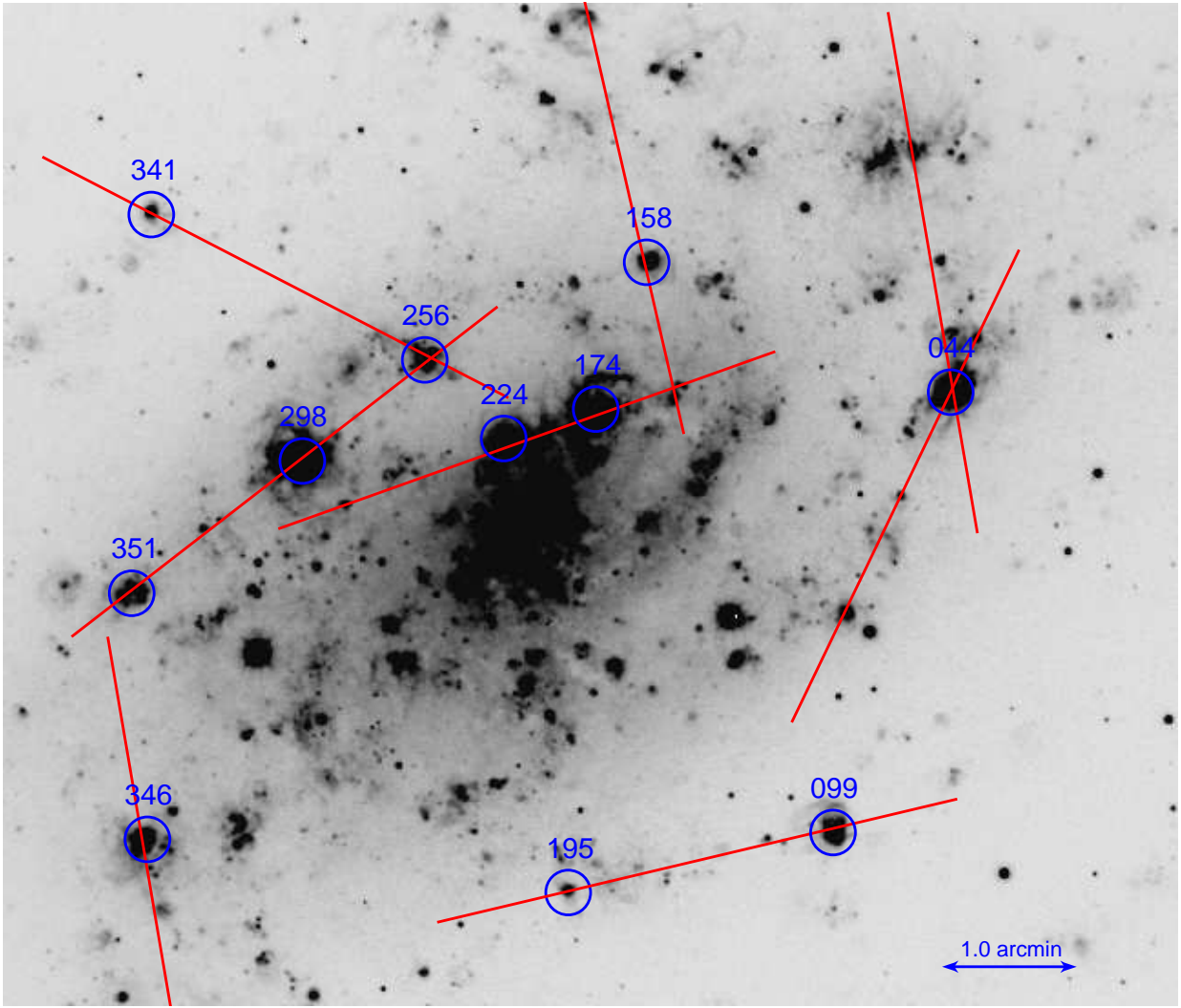}
\caption{Continuum-subtracted H$\alpha$ narrowband image of NGC~2403, taken with the 2.1 m telescope at Kitt Peak National Observatory. Blue circles enclose the \HII~regions studied in this paper, and the identification number of each \HII~region is quoted from \citet{1990A&A...237...23S}. The positions of long-slit spectroscopic observations are shown as red solid lines, which overlaid the \HII~regions. The scale in the bottom right corner indicates the 1$\arcmin$.0 length. North is at the up and east is to the left.} \label{Fig_img}
\end{figure*}

We selected 11 \HII~regions, bright and large enough for spatially resolved studies, inside NGC~2403 from the catalog of \citet[][hereafter denoted as S90]{1990A&A...237...23S} as the sample. Their identification numbers in S90 are 044, 099, 158, 174, 195, 224, 256, 298, 341, 346, and 351, respectively. Spectroscopic observations were taken during seven nights in the years 2007 and 2008, with the 2.16 m telescope mounted at XingLong station of National Astronomical Observatories of China \citep{2016PASP..128k5005F}, as part of a spectroscopic survey of \HII~regions in nearby galaxies \citep[][Lin et al. 2018, in preparation]{2014ChA&A..38..427K}. The OMR (Opto-Mechanics Research Inc.) long-slit spectrograph, equipped with a TEKTRONIX TEK1024 AR-coated back-illuminated CCD and a grating of 300 lines per millimeter (i.e., 200 \AA~mm$^{-1}$) blazed at 5500 \AA, were used to obtain spectra with a wavelength range of 3500$-$8000 \AA~and a spectral resolution of 500$-$550 (defined by $\lambda/\Delta\lambda$) at 5000 \AA. The length of the spectrographic slit is 4$\arcmin$. The slit width was adjusted to 2$\arcsec$.5 in accordance with the local seeing disk.

Exposure time for each \HII~region was $1800 \times 2$ or $1200 \times 3$ seconds. The slit was rotated to intersect as many \HII~regions as possible at every exposure. Figure \ref{Fig_img} shows spatial positions of the \HII~regions in NGC~2403 and the orientation of the slit during each observation.

Instrument bias and dome flats were recorded at the beginning and the end of every night. A He-Ar lamp was observed after every exposure of the objects for wavelength calibrations. Spectrophotometric standards were selected from the catalog of International Reference Stars \citep[IRS,][]{1991irsc.book.....C} and observed several times at every night for flux calibrations.

Table \ref{Tab_HIIs} lists basic properties of the \HII~regions and the information about the observations. We observed the No. 044 \HII~region twice with two different position angles of the slit. The two observations are identified as 044a and 044b, respectively, as shown in Table \ref{Tab_HIIs} and Figure \ref{Fig_ind}.

\begin{deluxetable*}{lccrcclrc}
\tabletypesize{\tiny}
\tablecaption{Basic Properties for the \HII~Regions
and Information About the Observations}
\tablewidth{0pc}
\tablehead{
\colhead{ID.\tablenotemark{a}} & \colhead{R.A.\tablenotemark{b}} & \colhead{Decl.\tablenotemark{b}} & \colhead{$F$(H$\alpha$)\tablenotemark{a,c}} & \colhead{Observation} & \colhead{Exposure} & \colhead{Standard} &
\colhead{Slit Angle\tablenotemark{d}} & \colhead{Airmass} \\
\colhead{} & \colhead{(J2000.0)} & \colhead{(J2000.0)} & \colhead{} & \colhead{Date} & \colhead{Time} & \colhead{Stars} & \colhead{} & \colhead{} \\
\colhead{(1)} & \colhead{(2)} & \colhead{(3)} & \colhead{(4)} & \colhead{(5)} & \colhead{(6)} & \colhead{(7)} & \colhead{(8)} & \colhead{(9)}
}
\startdata
044a & 07~36~20.0 & +65~37~07 & 24.442 \hspace*{0.3em} & 2008~Jan~01 & 1800~$\times$~2 & HD19445 & 9.2 \hspace*{1.2em} & 1.1064 \\
044b & 07~36~20.0 & +65~37~07 & 24.442 \hspace*{0.3em} & 2008~Jan~06 & 1800~$\times$~2 & HD19445 & $-$26.1 \hspace*{1.2em} & 1.1055 \\
099 & 07~36~28.8 & +65~33~51 & 5.460 \hspace*{0.3em} & 2008~Jan~04 & 1800~$\times$~2 & HE3 & $-$77.5 \hspace*{1.2em} & 1.1422 \\
158 & 07~36~41.8 & +65~38~06 & 2.689 \hspace*{0.3em} & 2008~Jan~29 & 1200~$\times$~3 & HILTNER600 & 13.0 \hspace*{1.2em} & 1.1524 \\
174 & 07~36~45.6 & +65~37~01 & 19.819 \hspace*{0.3em} & 2008~Jan~07 & 1800~$\times$~2 & G191B2B & $-$71.0 \hspace*{1.2em} & 1.1218 \\
195 & 07~36~48.0 & +65~33~26 & 0.317 \hspace*{0.3em} & 2008~Jan~04 & 1800~$\times$~2 & HE3 & $-$77.5 \hspace*{1.2em} & 1.1422 \\
244 & 07~36~52.2 & +65~36~21 & 9.878 \hspace*{0.3em} & 2008~Jan~07 & 1800~$\times$~2 & G191B2B & $-$71.0 \hspace*{1.2em} & 1.1218 \\
256 & 07~36~57.8 & +65~36~24 & 4.940 \hspace*{0.3em} & 2008~Jan~01 & 1200~$\times$~3 & HE3 & 63.0 \hspace*{1.2em} & 1.1124 \\
298 & 07~36~06.7 & +65~36~39 & 30.010 \hspace*{0.3em} & 2007~Jan~18 & 1800~$\times$~2 & G191B2B & $-$53.0 \hspace*{1.2em} & 1.1059 \\
341 & 07~37~17.5 & +65~38~29 & 1.786 \hspace*{0.3em} & 2008~Jan~01 & 1200~$\times$~3 & HE3 & 63.0 \hspace*{1.2em} & 1.1124 \\
346 & 07~37~18.1 & +65~33~50 & 3.049 \hspace*{0.3em} & 2008~Jan~01 & 1800~$\times$~2 & FEIGE34 & 9.2 \hspace*{1.2em} & 1.1588 \\
351 & 07~37~19.2 & +65~35~40 & 1.601 \hspace*{0.3em} & 2007~Jan~18 & 1800~$\times$~2 & G191B2B & $-$53.0 \hspace*{1.2em} & 1.1059
\enddata
\tablecomments{\scriptsize Columns: (1) Identification numbers of the \HII~regions; (2) Right ascension in the format of hour~minute~second; (3) Declination in the format of degree~minute~second; (4) H$\alpha$ flux in units of $10^{-13}~\mathrm{ergs~s^{-1}~cm^{-2}}$; (5) Observation dates in the format of year~month~date; (6) Exposure time in units of seconds; (7) Standard stars selected from the IRS catalog \citep{1991irsc.book.....C}; (8) Rotation angle of the spectrographic slit in units of degrees; (9) Airmass at each slit position.}
\tablenotetext{a}{\scriptsize ~Data are obtained from \citet{1990A&A...237...23S}.}
\tablenotetext{b}{\scriptsize ~Data are obtained from astrometry information in the continuum-subtracted H$\alpha$ narrowband image of NGC~2403, taken with the 2.1 m telescope at Kitt Peak National Observatory.}
\tablenotetext{c}{\scriptsize ~The $F$(H$\alpha$) for the No. 351 \HII~region in this table is in reality a sum of fluxes for the \HII~regions Nos. 348, 351, and 352, since the three \HII~regions are unable to be resolved in our observations.}
\tablenotetext{d}{\scriptsize ~The orientation from north to east is defined to be positive, and that from north to west is defined to be negative.}
\label{Tab_HIIs}
\end{deluxetable*}

\section{\textbf{DATA REDUCTION AND MEASUREMENTS}}\label{Sec_data}

\begin{figure*}[!ht]
\centering
\includegraphics[width=2.0\columnwidth]{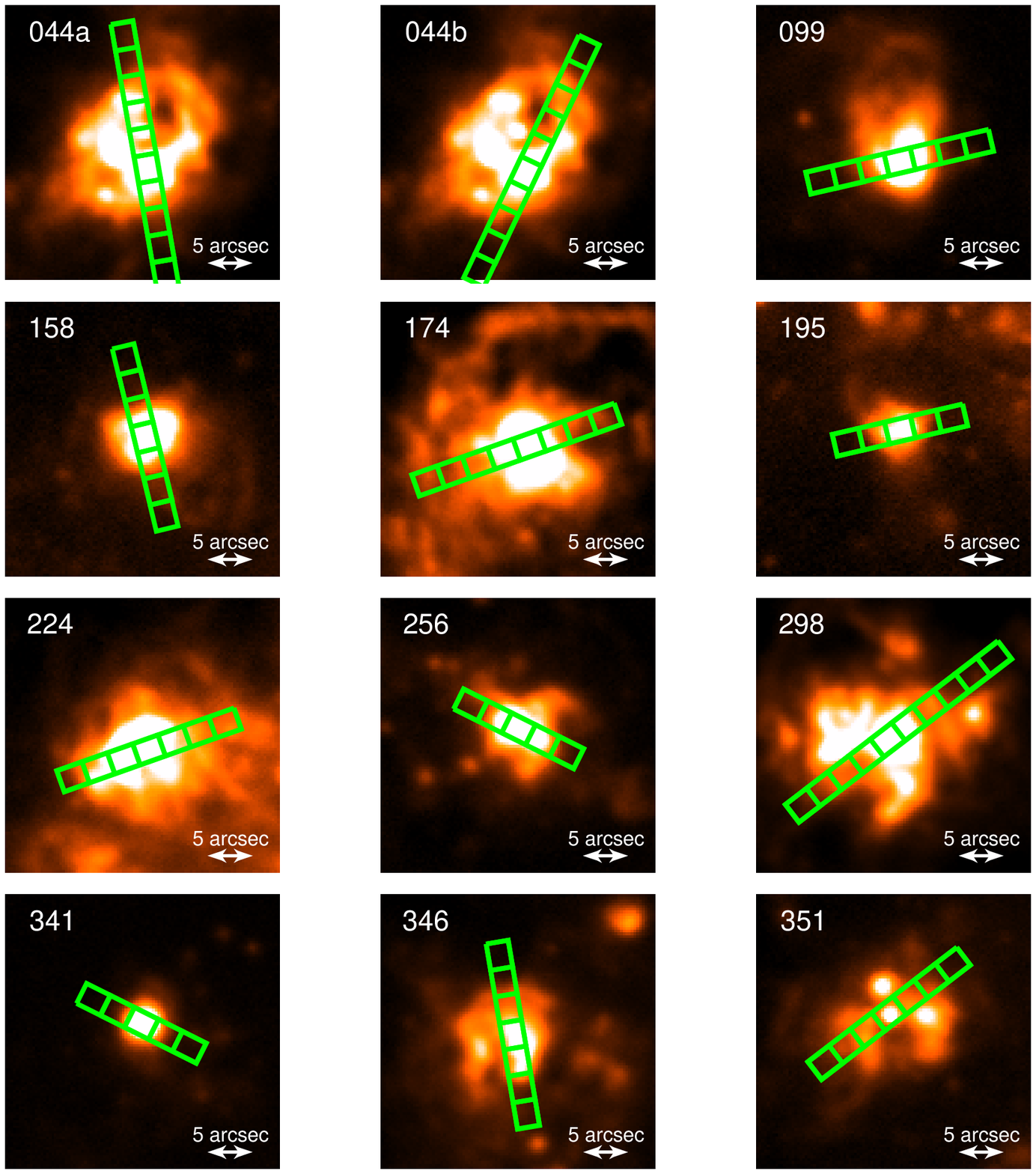}
\caption{Continuum-subtracted H$\alpha$ narrowband images of the \HII~regions studied in this paper, intercepted from the image displayed in Figure \ref{Fig_img}. Green boxes represent rectangular apertures used for extracting spectra on the slit. In each panel, the identification number of the \HII~region is displayed in the top left corner; the spatial scale in the bottom right corner indicates the 5$\arcsec$ length; north is at the up and east is to the left.} \label{Fig_ind}
\end{figure*}

The raw data were reduced by using the IRAF software.\footnote{IRAF (the Image Reduction and Analysis Facility) is a general purpose software system for the reduction and analysis of astronomical data, and is distributed by the National Optical Astronomy Observatories, which is operated by the Association of Universities for Research in Astronomy (AURA), Inc., under cooperative agreement with the National Science Foundation.} After conventional steps of data reduction including bias subtraction, flat-field correction, and cosmic-ray removal (details of the processes will be presented in Lin et al. 2018, in preparation), we extract spectra radially for each \HII~region by employing rectangular apertures on the slit. The size of the apertures is set to $3\arcsec.0 \times 2\arcsec.5$, corresponding to a physical scale of $66 \times 55~\mathrm{pc}^2$, approximately, with the distance of 3.5 Mpc adopted for NGC~2403 \citep{2006MNRAS.369.1780V}. During the radial spectra extraction, the first aperture was placed at the position of the H$\alpha$ emission peak in a nebula and defined as the central aperture; other apertures were placed along the slit at both sides of the central aperture with no gap between every two adjacent apertures; the number of the apertures applied to a nebula depends on the extension scale of H$\alpha$ emission along the spatial axis in the spatial-dispersion plane of the raw data; the outmost aperture at each side was placed by visually determining the outskirts of the H$\alpha$ profile through the interactive window of IRAF.\footnote{In this case, the outmost apertures for the \HII~regions do not stand at the same H$\alpha$ intensity level.} For each \HII~region, we also employed a large aperture integrating all of the 3 arcsec length apertures for extracting an integrated spectrum. Figure \ref{Fig_ind} illustrates the positions of the apertures placed on the \HII~regions studied in our work. During the extraction of spectra from the apertures, we traced the trajectories along the dispersion axis, by applying a common function to all apertures in the same \HII~region. Continuum points along the trace were obtained by summing enough dispersion lines and sampled to fit the tracing function manually under the IRAF interactive mode. Background levels were estimated from blank areas on the slit and subtracted from the object spectra.

\begin{figure*}[!ht]
\centering
\vspace*{-10mm}
\hspace*{-34mm}
\includegraphics[width=3.0\columnwidth]{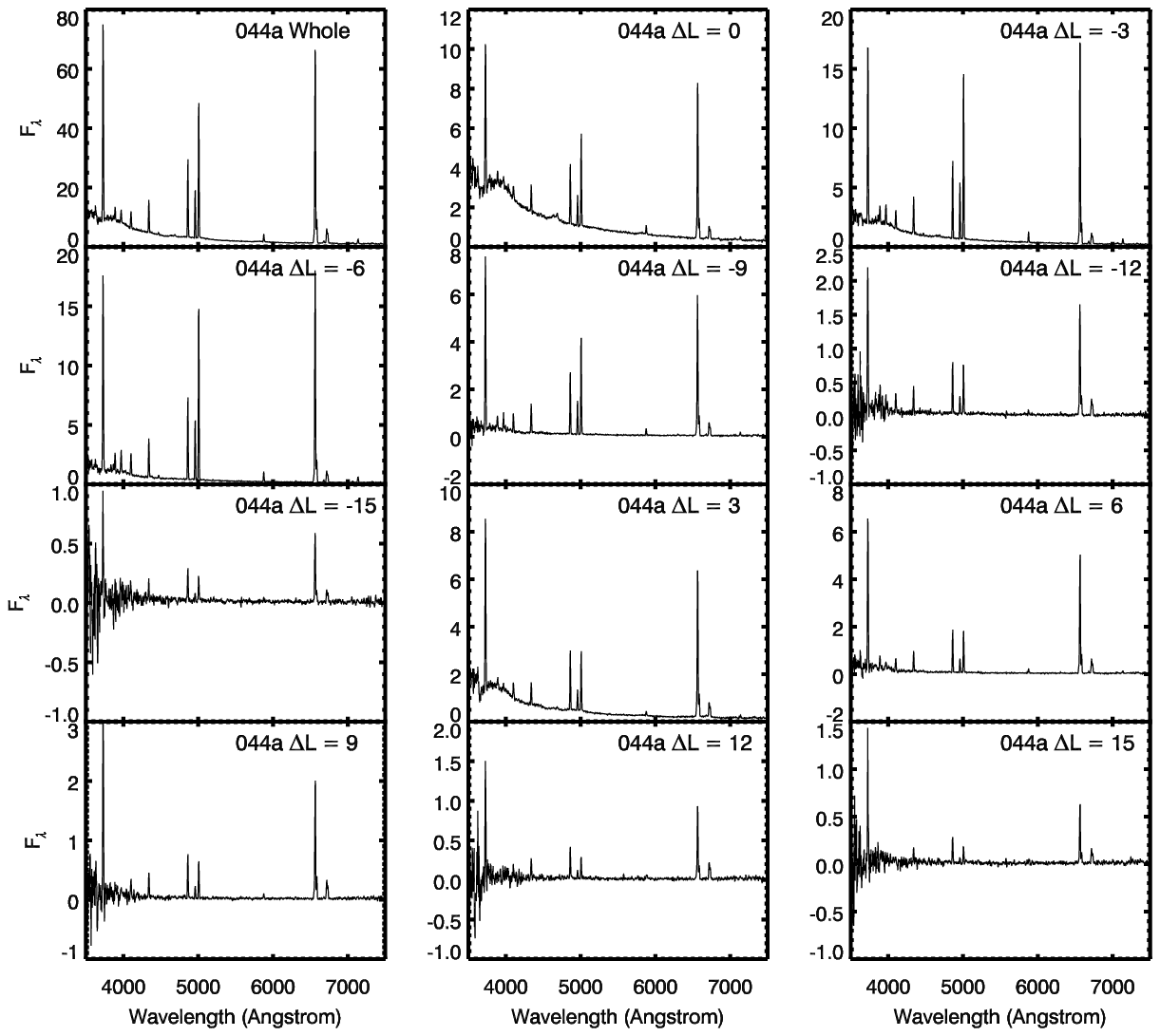}
\vspace*{-5mm}
\caption{Observed spectra for all spectrophotometric units in the No. 044a \HII~region as an example to convey the quality of the data obtained by our observations. The flux ($F_\mathrm{\lambda}$) in each spectrum is in units of  $10^{-15}~\mathrm{ergs~s^{-1}~cm^{-2}~\AA^{-1}}$. $\Delta$L in each diagram identifies the spatial coordinate in units of arcsecond, where $\Delta$L = 0 represents the center of the \HII~region defined in Section \ref{Sec_data}; the top left panel shows the spectra extracted from a large aperture integrating all the 3 arcsec length apertures.} \label{Fig_spec}
\end{figure*}

The dispersion of the extracted spectra was calibrated to wavelength by comparing with the spectra of the He-Ar lamp. Flux calibrations were performed by adopting the spectra of the standard stars and the atmospheric extinction curve at Xinglong station. All the spectra were corrected for Galactic foreground extinction, with the \citet{1989ApJ...345..245C} extinction curve and the total-to-selective extinction ratio $R_\mathrm{V} = 3.1$ is utilized. The color excess of Galactic extinction $E(\mathrm{B}-\mathrm{V})_{\mathrm{GAL}} = 0.04$ mag for NGC~2403 was obtained from the \citet{1998ApJ...500..525S} Galactic dust map. Calibrated spectra for the No. 044a \HII~region are displayed in Figure \ref{Fig_spec} as a representative example of our spectral data.

We measured fluxes of the emission lines $\OII\lambda3727$, H$\gamma$, H$\beta$, $\OIII\lambda\lambda4959,5007$, H$\alpha$, and $\NII\lambda6583$, by fitting Gaussian profiles through the MPFITEXPR algorithm \citep{2009ASPC..411..251M}. A major part of uncertainties in the fluxes were calculated by using the expression addressed in \citet{2002MNRAS.329..315C, 2009ApJ...700..309B}: $\sigma_\mathrm{line} = \sigma_\mathrm{con}N_\mathrm{pix}^{1/2}[1+EW_\mathrm{line}/(N_\mathrm{pix}\Delta_\mathrm{\lambda})]^{1/2}$, where $\sigma_\mathrm{line}$ is the error in the flux of the emission line, $\sigma_\mathrm{con}$ is the standard deviation in the continuum near the emission line, $N_\mathrm{pix}$ is the number of pixels sampled in the measurements of the emission line, $EW_\mathrm{line}$ is the equivalent width of the emission line, and $\Delta_\mathrm{\lambda}$ is the minimum wavelength-interval (in units of \AA~per pixel) of the spectral data. Errors in the Gaussian fitting were also combined into final uncertainties in the emission-line fluxes.

\begin{figure*}[!ht]
\centering
\vspace*{-12mm}
\hspace*{-48mm}
\includegraphics[width=3.3\columnwidth]{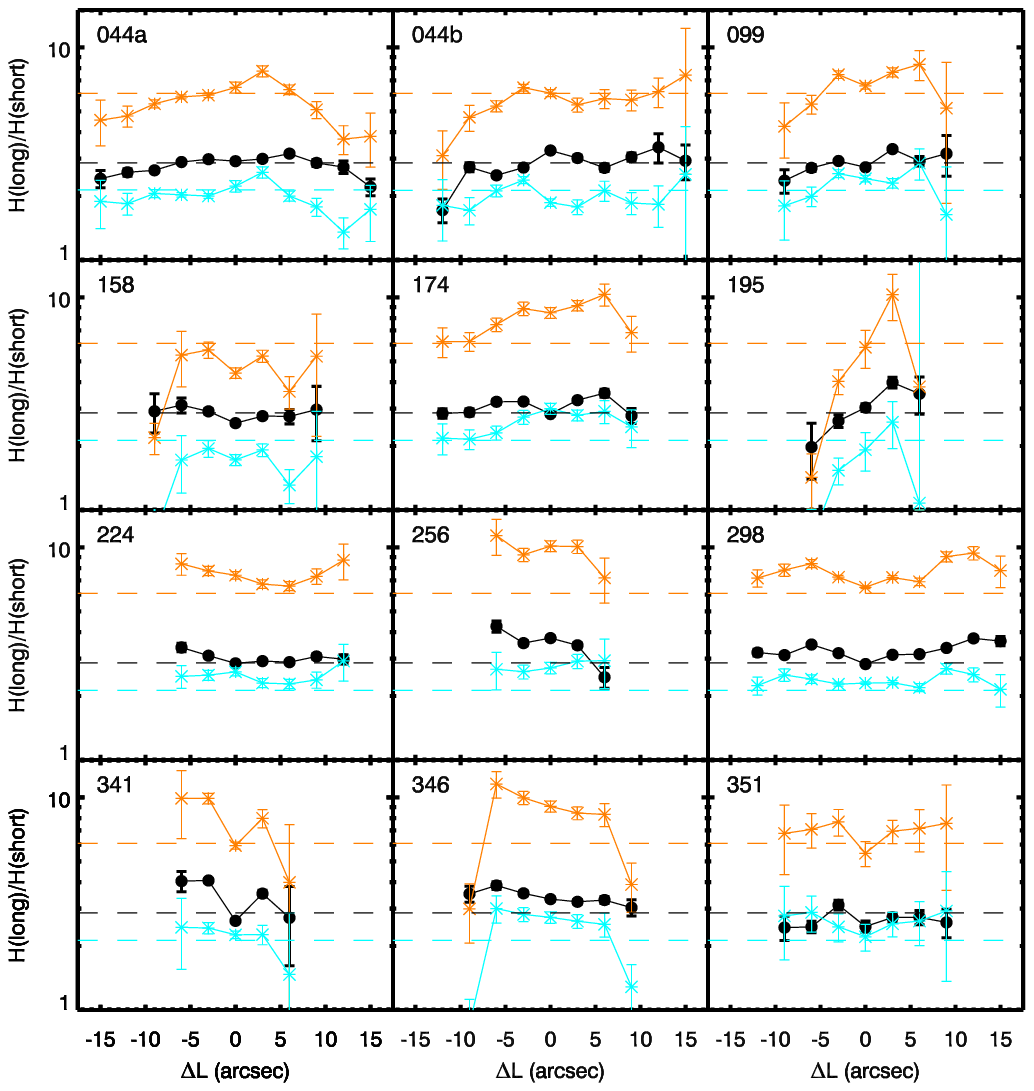}
\vspace*{-8mm}
\caption{Radial profiles of the observed flux ratios of H$\alpha$ to H$\beta$ (black filled circles), H$\alpha$ to H$\gamma$ (orange asterisks), and H$\beta$ to H$\gamma$ (cyan asterisks) in the individual \HII~regions. Dashed lines mark the theoretically assumed lower limits for H$\alpha$/H$\beta$ (= 2.86, black), H$\alpha$/H$\gamma$ (= 6.09, orange), and H$\beta$/H$\gamma$ (= 2.13, cyan). In each panel, the abscissa is the spatial coordinate in units of arcseconds, where $\Delta$L = 0 represents the center of one \HII~region defined in Section \ref{Sec_data}.} \label{Fig_Habc}
\end{figure*}

All the fluxes were then corrected for internal dust attenuation of NGC~2403, by means of the Balmer decrement H$\alpha$/H$\beta$ \citep{2006agna.book.....O}, on assumption of the \citet{1999PASP..111...63F} attenuation/extinction curve,\footnote{The terminology "\emph{attenuation}", strictly speaking, is not equivalent to "\emph{extinction}" \citep[see][for more details about the terminologies]{2001PASP..113.1449C, 2014ApJ...789...76M}. In this work , we are faced with dust "\emph{attenuation}" when studying the extraGalactic \HII~regions, whereas \citet{1999PASP..111...63F} has legislated dust "\emph{extinction}" as a function of wavelength by compiling a sample of Galactic point sources. However, we are still allowed to assume the wavelength-dependent attenuation for the \HII~regions in this paper parameterized by the \citet{1999PASP..111...63F} curve even though it was developed initially to depict "\emph{extinction}". Hereafter, we keep using the wording "\emph{attenuation}" in this paper albeit in some cases we may actually deal with \emph{extinction}.} the total-to-selective attenuation ratio $R_\mathrm{V} = 3.1$, and the intrinsic H$\alpha$-to-H$\beta$ ratio equal to 2.86 \citep{1987MNRAS.224..801H}. Figure \ref{Fig_Habc} shows radial profiles of the observed flux ratios H$\alpha$/H$\beta$, H$\alpha$/H$\gamma$, and H$\beta$/H$\gamma$ in all the \HII~regions studied in this paper. A generally uniform trend of the three Balmer decrements can be seen. The data with observed $\mathrm{H}\alpha/\mathrm{H}\beta \leq 2.86$ were considered with no internal dust attenuation, and we did not perform the attenuation correction in this case. Application of a different attenuation curve will cause slight changes in final results, which will be inspected in the Appendix.

\begin{figure}[!ht]
\centering
\hspace*{-10mm}
\includegraphics[width=1.33\columnwidth]{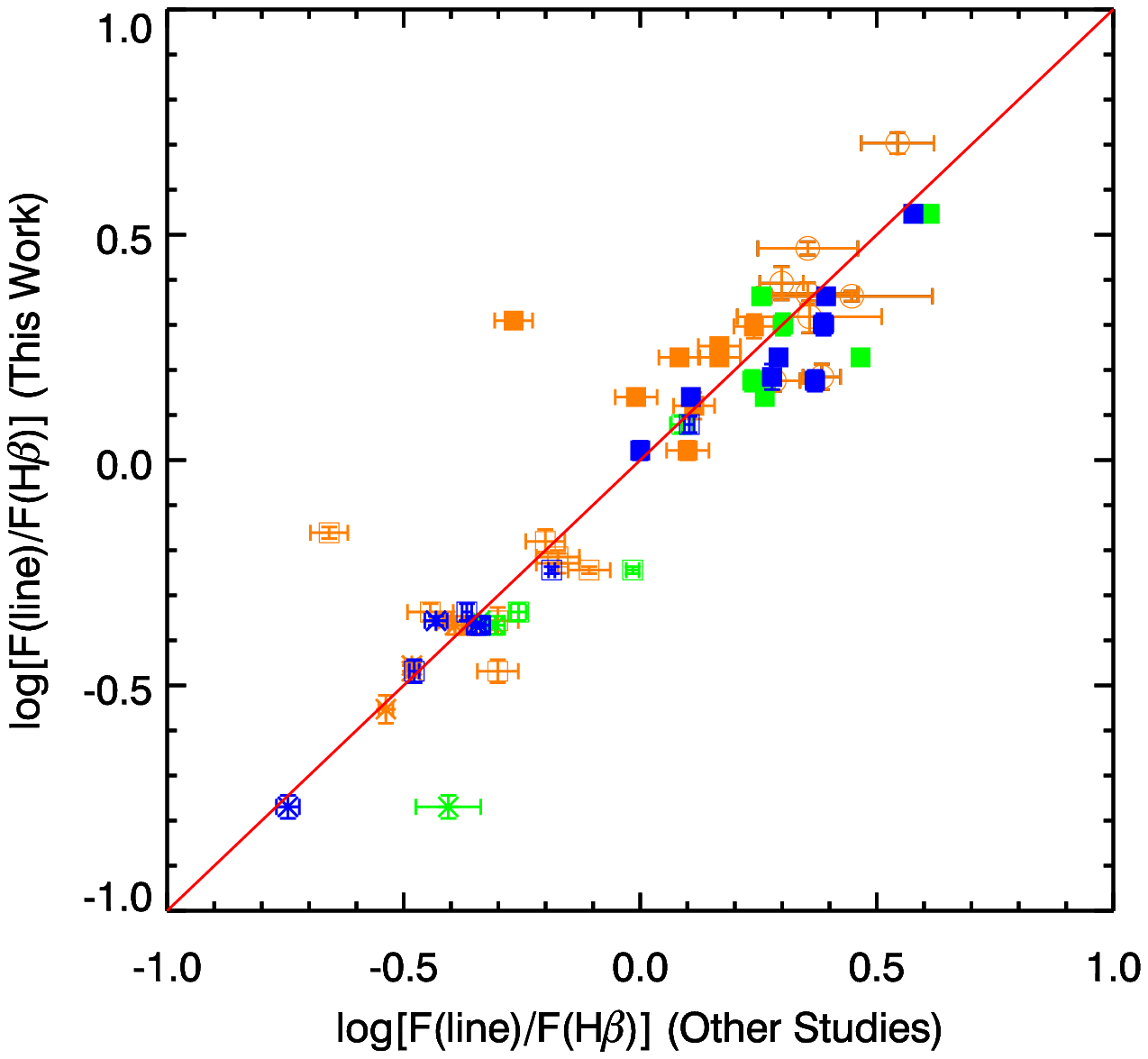}
\caption{Comparison of attenuation-corrected emission-line fluxes (relative to H$\beta$ flux) for the \HII~regions obtained through integrated measurements in our work with those in previous studies, including \citet[][in orange]{1997ApJ...489...63G}, \citet[][in green]{1999ApJ...513..168G}, and \citet[][in blue]{2013ApJ...775..128B}. The emission lines compared in this figure are $\OII\lambda3727$ (open circles), $\OIII\lambda4959$ (open squares), $\OIII\lambda5007$ (filled squares), and $\NII\lambda6583$ (asterisks). The line of unity is also shown as the red solid line.} \label{Fig_comp}
\end{figure}

\begin{figure*}[!ht]
\centering
\vspace*{-12mm}
\hspace*{-48mm}
\includegraphics[width=3.3\columnwidth]{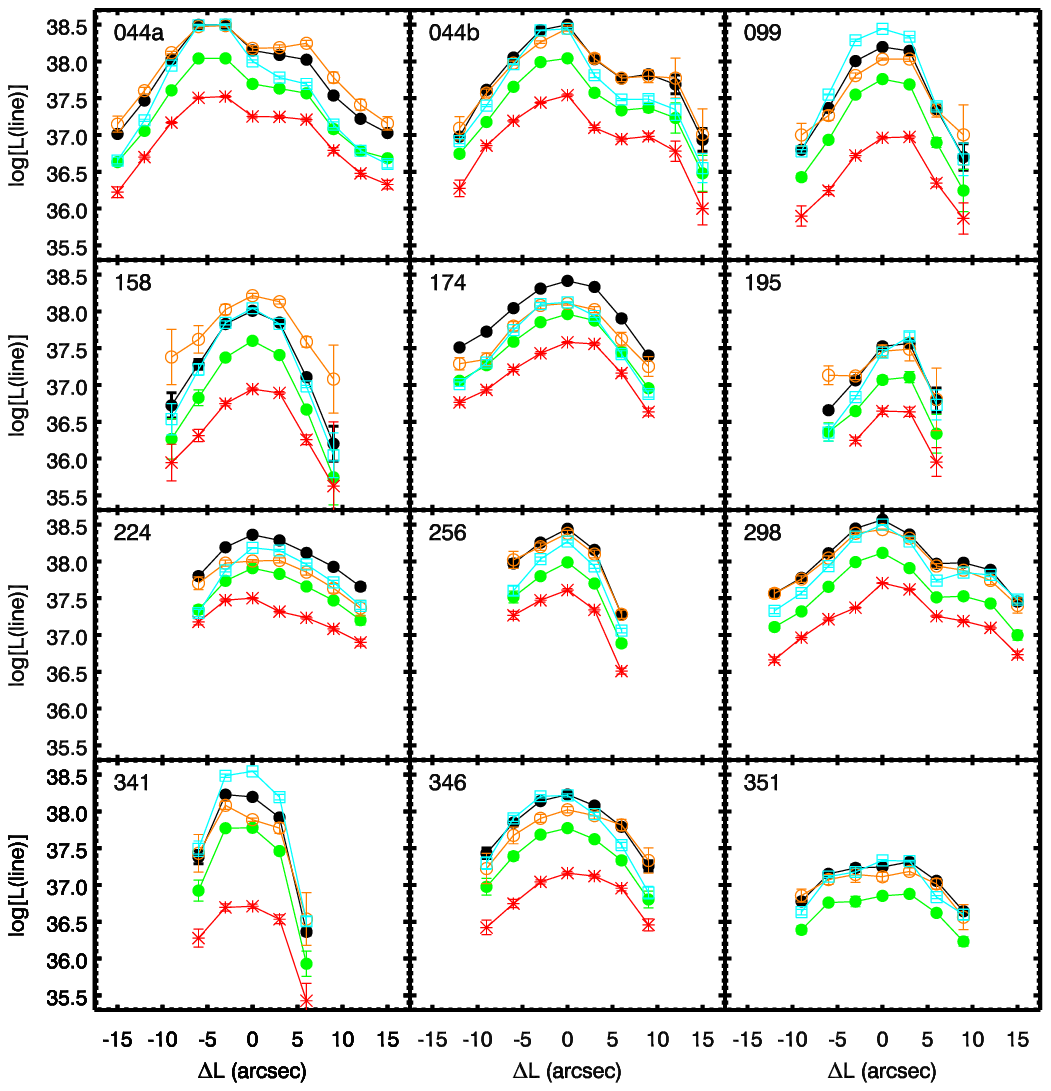}
\vspace*{-8mm}
\caption{Radial profiles of attenuation-corrected luminosities in the form of logarithm for the emission lines, $\OII\lambda3727$ (orange open circles), H$\beta$ (green filled circles), $\OIII\lambda\lambda4959,5007$ (cyan open squares), H$\alpha$ (black filled circles), and $\NII\lambda6583$ (red asterisks), in the individual \HII~regions. All the luminosities are in units of $\mathrm{ergs~s^{-1}}$) in the form of logarithm. In each panel, the abscissa is the spatial coordinate in units of arcsecond, where $\Delta$L = 0 represents the center of one \HII~region defined in Section \ref{Sec_data}.} \label{Fig_lgL0}
\end{figure*}

Table \ref{Tab_Flux} presents the observed fluxes of the emission lines (relative to H$\beta$) for all the Spectrophotometric Units, and Table \ref{Tab_Flux_0} presents the attenuation-corrected ones in the same form. We compared the attenuation-corrected flux ratios of $\OII\lambda3727$, $\OIII\lambda4959$, $\OIII\lambda5007$, and $\NII\lambda6583$, respectively, to H$\beta$ obtained through integrated measurements in our work, with those in \citet{1997ApJ...489...63G, 1999ApJ...513..168G, 2013ApJ...775..128B}, for the \HII~regions commonly sampled in these studies. As depicted in Figure \ref{Fig_comp}, the consistency between the normalized fluxes worked out by the different observations and measurements verifies the reliability of our work. Figure \ref{Fig_lgL0} shows radial profiles of attenuation-corrected luminosities in the form of logarithm for the emission lines, $\OII\lambda3727$, H$\beta$, $\OIII\lambda\lambda4959,5007$, H$\alpha$, and $\NII\lambda6583$ in the \HII~regions. The distributions of the luminosities are similar in general trend but distinct in detailed gradient for the different emission lines in each panel of the figure. The disparity between the gradients is likely to correlate with different sensitivities of these emission lines to the degree of ionization, which will be discussed in Section \ref{Sec_disc}.

\begin{deluxetable*}{lrrrrrrr}
\tabletypesize{\tiny}
\tablecaption{Observed Emission-line Fluxes for the Spectrophotometric Units}
\tablewidth{0pc}
\tablehead{
\colhead{ID.} & \colhead{H$\beta$} & \colhead{$\OII\lambda3727$/H$\beta$} & \colhead{H$\gamma$/H$\beta$} & \colhead{$\OIII\lambda4959$/H$\beta$} & \colhead{$\OIII\lambda5007$/H$\beta$} & \colhead{H$\alpha$/H$\beta$} & \colhead{$\NII\lambda6583$/H$\beta$}
}
\startdata
  044a\_0 & 33.76~$\pm$~0.66 & 3.00~$\pm$~0.16 & 0.45~$\pm$~0.03 & 0.49~$\pm$~0.02 & 1.55~$\pm$~0.04 & 2.92~$\pm$~0.06 & 0.37~$\pm$~0.01 \\
 044a\_-3 & 70.74~$\pm$~0.69 & 2.69~$\pm$~0.08 & 0.50~$\pm$~0.01 & 0.72~$\pm$~0.01 & 2.18~$\pm$~0.02 & 2.98~$\pm$~0.03 & 0.32~$\pm$~0.01 \\
 044a\_-6 & 77.01~$\pm$~0.68 & 2.69~$\pm$~0.07 & 0.49~$\pm$~0.01 & 0.71~$\pm$~0.01 & 2.13~$\pm$~0.02 & 2.89~$\pm$~0.03 & 0.29~$\pm$~0.01 \\
 044a\_-9 & 29.23~$\pm$~0.43 & 3.26~$\pm$~0.13 & 0.49~$\pm$~0.02 & 0.55~$\pm$~0.01 & 1.62~$\pm$~0.03 & 2.64~$\pm$~0.04 & 0.36~$\pm$~0.01 \\
044a\_-12 & 8.21~$\pm$~0.30 & 3.52~$\pm$~0.39 & 0.54~$\pm$~0.06 & 0.37~$\pm$~0.03 & 1.03~$\pm$~0.05 & 2.58~$\pm$~0.10 & 0.44~$\pm$~0.03 \\
044a\_-15 & 3.09~$\pm$~0.27 & 3.28~$\pm$~0.86 & 0.53~$\pm$~0.14 & 0.26~$\pm$~0.08 & 0.79~$\pm$~0.11 & 2.41~$\pm$~0.23 & 0.39~$\pm$~0.07 \\
  044a\_3 & 27.22~$\pm$~0.50 & 3.46~$\pm$~0.17 & 0.39~$\pm$~0.02 & 0.36~$\pm$~0.02 & 1.05~$\pm$~0.03 & 2.99~$\pm$~0.06 & 0.43~$\pm$~0.01 \\
  044a\_6 & 19.93~$\pm$~0.39 & 4.40~$\pm$~0.21 & 0.50~$\pm$~0.03 & 0.35~$\pm$~0.02 & 1.01~$\pm$~0.03 & 3.16~$\pm$~0.07 & 0.49~$\pm$~0.02 \\
  044a\_9 & 8.67~$\pm$~0.31 & 5.02~$\pm$~0.47 & 0.56~$\pm$~0.05 & 0.29~$\pm$~0.03 & 0.86~$\pm$~0.04 & 2.86~$\pm$~0.11 & 0.51~$\pm$~0.03 \\
 044a\_12 & 4.38~$\pm$~0.28 & 4.24~$\pm$~0.76 & 0.74~$\pm$~0.12 & 0.28~$\pm$~0.06 & 0.73~$\pm$~0.07 & 2.74~$\pm$~0.19 & 0.50~$\pm$~0.06 \\
 044a\_15 & 3.48~$\pm$~0.29 & 2.98~$\pm$~0.68 & 0.58~$\pm$~0.17 & 0.22~$\pm$~0.07 & 0.61~$\pm$~0.08 & 2.21~$\pm$~0.20 & 0.44~$\pm$~0.07 \\
 044a\_wh & 289.11~$\pm$~1.72 & 2.93~$\pm$~0.05 & 0.43~$\pm$~0.01 & 0.61~$\pm$~0.01 & 1.79~$\pm$~0.01 & 2.88~$\pm$~0.02 & 0.35~$\pm$~0.00 \\
\nodata & & & & & & &
\enddata
\tablecomments{\scriptsize The fluxes in this table are  in units of $10^{-15}~\mathrm{ergs~s^{-1}~cm^{-2}}$ and not corrected for internal dust attenuation. This table is available in its entirety in the online journal. A portion is shown here for guidance regarding its form and content.}
\label{Tab_Flux}
\end{deluxetable*}

It is worth pointing out that, in spectroscopic observations, apparent positions of observed objects are likely to be shifted by atmospheric refraction. The displacements of apparent from true positions will increase at larger airmasses or through shorter wavelength channels, or with a position-angle closer to 90$^{\circ}$ for slit-spectrographs specifically. Notwithstanding, in our work, given that all the objects were observed near the zenith, at the airmasses $\sim 1.11 - 1.15$ as shown in Table \ref{Tab_HIIs}, the impacts of atmospheric refraction are supposed to be trivial. In this case, during the process of data reduction, we did not correct the effects of the differential atmospheric refraction. Readers are referred to \citet{1982PASP...94..715F} for a comprehensive introduction about atmospheric refraction and its influences on slit spectrometry.

\section{\textbf{RESULTS}}\label{Sec_res}

With the reduction and the measurements of the data described above, we derive oxygen abundances for each of the spectrophotometric units in the \HII~regions from the four widely used strong-line indices, $R23$, $N2O2$, $N2$, and $O3N2$. These indices have been diversely calibrated to oxygen abundance, through empirical fits of not only observed relationships between $T_\mathrm{e}$ and the indices but also theoretical grids from photoionization models. The calibrations adopted in our work are listed as follows.

The $R23$ calibration addressed in \citet[][hereafter denoted as $R23_\mathrm{K99}$]{1999ApJ...514..544K},
\begin{equation}
\begin{aligned}
12 + \log(\mathrm{O/H})_\mathrm{low} = 12.0 - 4.944 + 0.767 x + 0.602 x^2 \\
- y (0.29 + 0.332 x - 0.331 x^2) ,
\label{Eq_R23l_K99}
\end{aligned}
\end{equation}
\begin{equation}
\begin{aligned}
12 + \log(\mathrm{O/H})_\mathrm{high} = 12.0 - 2.939 - 0.2 x - 0.237 x^2 \\
- 0.305 x^3 - 0.0283 x^4 \\
- y (0.0047 - 0.0221 x - 0.102 x^2 - 0.0817 x^3 - 0.00717 x^4) ,
\label{Eq_R23h_K99}
\end{aligned}
\end{equation}
where $x = \log[(\OII\lambda3727+\OIII\lambda\lambda4959,5007)/\mathrm{H}\beta]$, $y = \log(\OIII\lambda\lambda4959,5007/\OII\lambda3727)$;

The $R23$ calibration addressed in \citet[][hereafter denoted as $R23_\mathrm{P05}$]{2005ApJ...631..231P},
\begin{equation}
 12 + \log(\mathrm{O/H})_\mathrm{low} = \frac{x + 106.4 + 106.8 y - 3.40 y^2}{17.72 + 6.60 y + 6.95 y^2 - 0.302 x} ,
\label{Eq_R23l_P05}
\end{equation}
\begin{equation}
 12 + \log(\mathrm{O/H})_\mathrm{high} = \frac{x + 726.1 + 842.2 y + 337.5 y^2}{85.96 + 82.76 y + 43.98 y^2 + 1.793 x} ,
\label{Eq_R23h_P05}
\end{equation}
where $x = (\OII\lambda3727+\OIII\lambda\lambda4959,5007)/\mathrm{H}\beta$, $y = \OIII\lambda\lambda4959,5007/(\OII\lambda3727+\OIII\lambda\lambda4959,5007)$;

The $N2O2$ calibration addressed in \citet[][hereafter denoted as $N2O2_\mathrm{K02}$]{2002ApJS..142...35K},
\begin{equation}
 12 + \log(\mathrm{O/H}) = \log(1.54020 + 1.26602 x + 0.167977 x^2) + 8.93 ,
\label{Eq_N2O2_K02}
\end{equation}
where $x = \log(\NII\lambda6583/\OII\lambda\lambda3727)$;

The $N2O2$ calibration addressed in \citet[][hereafter denoted as $N2O2_\mathrm{B07}$]{2007ApJ...656..186B},
\begin{equation}
 12 + \log(\mathrm{O/H}) = 8.66 + 0.36 x - 0.17 x^2 ,
\label{Eq_N2O2_B07}
\end{equation}
where $x = \log(\NII\lambda6583/\OII\lambda\lambda3727)$;

The $N2$ calibration addressed in \citet[][hereafter denoted as $N2_\mathrm{P04}$]{2004MNRAS.348L..59P},
\begin{equation}
 12 + \log(\mathrm{O/H}) = 9.37 + 2.03 x + 1.26 x^2 + 0.32 x^3 ,
\label{Eq_N2_P04}
\end{equation}
where $x = \log(\NII\lambda6583/\mathrm{H}\alpha)$;

The $O3N2$ calibration addressed in \citet[][hereafter denoted as $O3N2_\mathrm{P04}$]{2004MNRAS.348L..59P},
\begin{equation}
 12 + \log(\mathrm{O/H}) = 8.73 - 0.32 x ,
\label{Eq_O3N2_P04}
\end{equation}
where $x = \log[(\OIII\lambda5007/\mathrm{H}\beta)/(\NII\lambda6583/\mathrm{H}\alpha)]$.

Since the $R23$ diagnostics suffer from a double-value problem (i.e., a fixed value for $R23$ corresponds to both a low value and a high value for $12 + \log(\mathrm{O/H})$), we choose the upper branch of the $R23$ diagnostics throughout the work, except for the No. 158 \HII~region behaving with apparent features of low metallicity to which we apply the lower branch, so as to avoid the degeneracy, in accordance with our existing knowledge cognizing NGC~2403 as a metal-rich galaxy. There are in total six sets of empirical oxygen abundances obtained for each nebula in our work. Among all the calibrations listed above, $R23_\mathrm{P05}$, $N2O2_\mathrm{B07}$, $N2_\mathrm{P04}$, and $O3N2_\mathrm{P04}$ are developed through observational approaches, while $R23_\mathrm{K99}$ and $N2O2_\mathrm{K02}$ are formulated theoretically on the basis of photoionization models. Some estimates from $R23$ or $N2O2$ are defined to be invalid. The invalidity of $R23$ occurs when the upper branch of the $R23$ diagnostics results in low abundances (or the lower branch leads to high values for the No. 158 \HII~region);\footnote{We are unable to define a criterion of the single index $R23$ to select "valid" data prior to estimating oxygen abundances, because the $R23$ diagnostics involve a combinative effect of the two indices $R23$ and $\log(\OIII\lambda\lambda4959,5007/\OII\lambda3727)$. The examination of the resulting values is an optimal way at this stage.} at the same time, estimates from $N2O2$ are picked out of valid results if $N2O2 \leq 1.2$, since the calibrations of $N2O2$ were carried out with high-metallicity samples, or high-metallicity zones in model grids, and applicable to $N2O2 > 1.2$ for reliable estimates \citep{2002ApJS..142...35K, 2007ApJ...656..186B}.

\begin{figure*}[!ht]
\centering
\vspace*{-12mm}
\hspace*{-48mm}
\includegraphics[width=3.3\columnwidth]{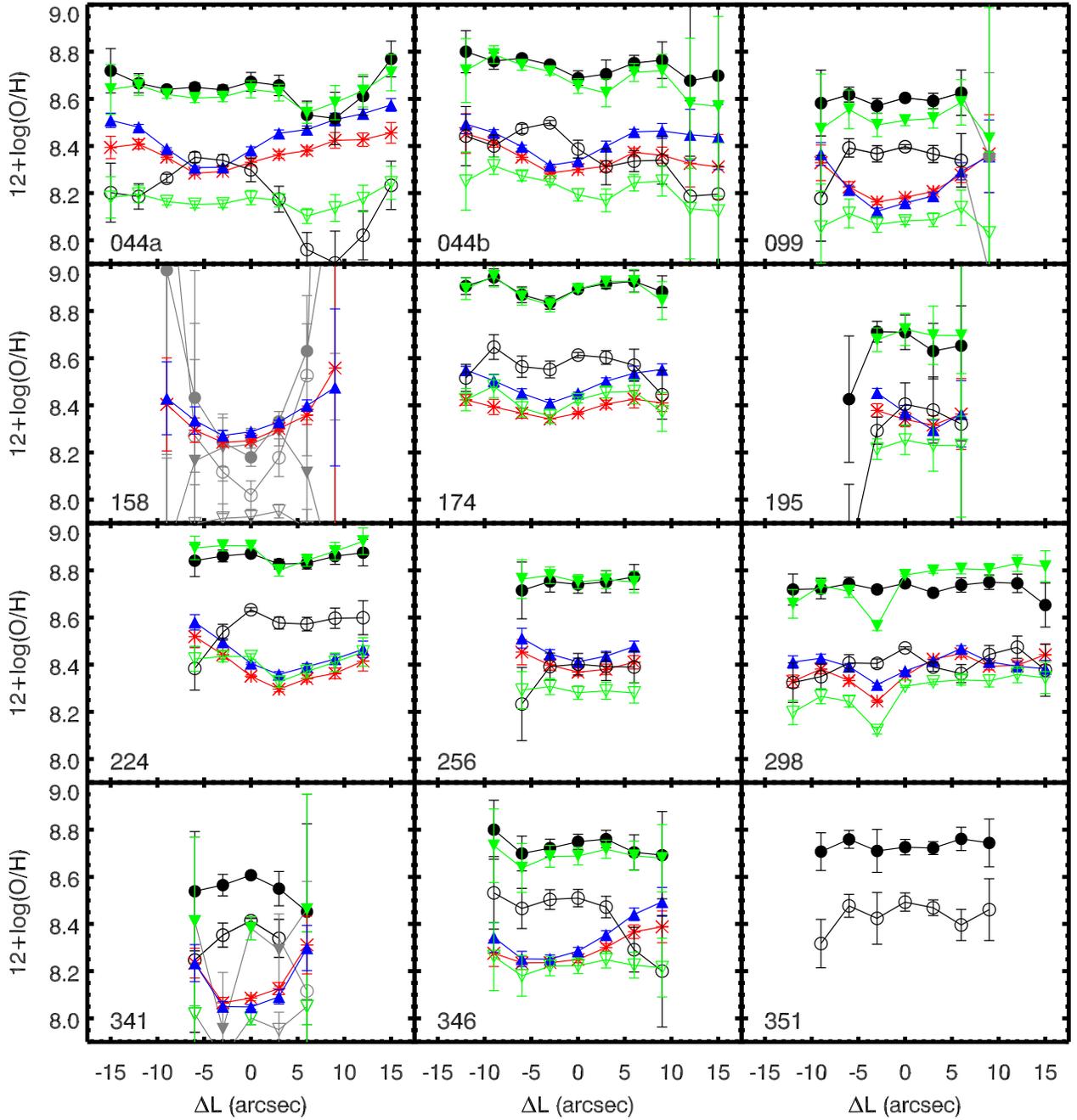}
\vspace*{-8mm}
\caption{Radial profiles of the oxygen abundances derived from $R23_\mathrm{K99}$ (black filled circles), $R23_\mathrm{P05}$ (black open circles), $N2_\mathrm{P04}$ (red asterisks), $O3N2_\mathrm{P04}$ (blue filled triangles), $N2O2_\mathrm{K02}$ (green filled upside-down triangles), and $N2O2_\mathrm{B07}$ (green open upside-down triangles) in the individual \HII~regions. The gray color marks the data out of the calibration ranges, considered to be invalid results as explained in Section \ref{Sec_res}. In each panel, the abscissa is the spatial coordinate in units of arcseconds, where $\Delta$L = 0 represents the center of one \HII~region defined in Section \ref{Sec_data}.} \label{Fig_abun}
\end{figure*}

Figure \ref{Fig_abun} shows radial profiles of these "strong-line" abundances in the \HII~regions. The invalid estimates from $R23$ or $N2O2$ are marked in gray in the diagrams, including the rightmost $R23$ data points from the No. 099 \HII~region, all the $R23$ and $N2O2$ data points from the No. 158 \HII~region, the rightmost $R23_\mathrm{P05}$ data point and the $\Delta\mathrm{L} = \pm3$ $N2O2$ data points from the No. 341 \HII~region. Also in this figure, we do not plot the leftmost data points in the $N2O2$, $N2$, and $O3N2$ profiles for the No. 195 \HII~region as well as the whole $N2O2$, $N2$, and $O3N2$ profiles for the No. 351 \HII~region, because the $\NII\lambda6583$ emission lines were not well deblended from H$\alpha$ for these spectra in the flux measurements.

Systematic deviation between the diagnostics based on observations and models can be obviously seen from each of the diagrams in Figure \ref{Fig_abun}. For the same spectral index, the calibrations on the basis of photoionization models leads to higher oxygen abundances (O/H) than the observation-based calibrations (i.e., $R23_\mathrm{K99}$ higher than $R23_\mathrm{P05}$, and $N2O2_\mathrm{K02}$ higher than $N2O2_\mathrm{B07}$) by 0.2$-$0.5 dex. The offsets appear to be constant at different nebular radii. On the other hand, the diagnostics with different spectral indices but identically calibrated via the observational relations with $T_\mathrm{e}$ (i.e., $R23_\mathrm{P05}$, $N2O2_\mathrm{B07}$, $N2_\mathrm{P04}$, and $O3N2_\mathrm{P04}$) result in approximately comparable abundance levels. The excesses of the model-based "strong-line" abundances to the observation-based ones have also been found in many other studies with integrated measurements of \HII~regions \citep[e.g.,][]{2003ApJ...591..801K, 2010ApJ...720.1738P, 2016MNRAS.457.3678P} or galaxies as a whole \citep[e.g.,][]{2006ApJ...652..257L, 2008ApJ...681.1183K, 2012ApJ...750..120Z}.

A remarkable feature in Figure \ref{Fig_abun} is the similarity and disparity between radial distributions of these "strong-line" abundances. We can see from each diagram that, $R23_\mathrm{K99}$ and $N2O2_\mathrm{K02}$ track each other exactly; $N2O2_\mathrm{B07}$ follows $R23_\mathrm{K99}$ and $N2O2_\mathrm{K02}$ in the shape of the profile despite the offset;\footnote{The exception is the No. 341 \HII~region, where $N2O2 < -1.0$ and thus the estimates from $N2O2$ become unreliable in this range.} the upturns and the downtrends of $R23_\mathrm{P05}$ roughly correlate with those of $R23_\mathrm{K99}$, $N2O2_\mathrm{K02}$, and $N2O2_\mathrm{B07}$, but the fluctuation is more intensive, yielding a radial variation $> 0.2$ dex in most cases (even up to 0.4 dex for No. 044a \HII~region); the estimates derived from $N2$ and $O3N2$ present an apparent gradient, increasing from the center to the edge of each nebula by $\sim 0.2 - 0.3$ dex on average.\footnote{The exception is No. 298, which shows approximately flat abundance profiles in Figure \ref{Fig_abun}. The \emph{Hubble Space Telescope} has resolved multiple ionizing sources in this \HII~region \citep{1999AJ....117.1249D}.}

\begin{deluxetable*}{lrrrrrrr}
\tabletypesize{\tiny}
\tablecaption{Attenuation-corrected Emission-line Fluxes for the Spectrophotometric Units}
\tablewidth{0pc}
\tablehead{
\colhead{ID.} & \colhead{H$\beta$} & \colhead{$\OII\lambda3727$/H$\beta$} & \colhead{H$\gamma$/H$\beta$} & \colhead{$\OIII\lambda4959$/H$\beta$} & \colhead{$\OIII\lambda5007$/H$\beta$} & \colhead{H$\alpha$/H$\beta$} & \colhead{$\NII\lambda6583$/H$\beta$}
}
\startdata
  044a\_0 & 35.69~$\pm$~2.22 & 3.04~$\pm$~0.34 & 0.45~$\pm$~0.05 & 0.49~$\pm$~0.04 & 1.54~$\pm$~0.13 & 2.86~$\pm$~0.21 & 0.36~$\pm$~0.03 \\
 044a\_-3 & 79.56~$\pm$~2.55 & 2.78~$\pm$~0.16 & 0.51~$\pm$~0.03 & 0.71~$\pm$~0.03 & 2.17~$\pm$~0.09 & 2.86~$\pm$~0.11 & 0.30~$\pm$~0.01 \\
 044a\_-6 & 79.36~$\pm$~2.33 & 2.71~$\pm$~0.14 & 0.50~$\pm$~0.02 & 0.71~$\pm$~0.03 & 2.13~$\pm$~0.09 & 2.86~$\pm$~0.10 & 0.29~$\pm$~0.01 \\
 044a\_-9 & 29.23~$\pm$~0.43 & 3.26~$\pm$~0.13 & 0.49~$\pm$~0.02 & 0.55~$\pm$~0.01 & 1.62~$\pm$~0.03 & 2.64~$\pm$~0.04 & 0.36~$\pm$~0.01 \\
044a\_-12 & 8.21~$\pm$~0.30 & 3.52~$\pm$~0.39 & 0.54~$\pm$~0.06 & 0.37~$\pm$~0.03 & 1.03~$\pm$~0.05 & 2.58~$\pm$~0.10 & 0.44~$\pm$~0.03 \\
044a\_-15 & 3.09~$\pm$~0.27 & 3.28~$\pm$~0.86 & 0.53~$\pm$~0.14 & 0.26~$\pm$~0.08 & 0.79~$\pm$~0.11 & 2.41~$\pm$~0.23 & 0.39~$\pm$~0.07 \\
  044a\_3 & 30.94~$\pm$~1.87 & 3.60~$\pm$~0.38 & 0.39~$\pm$~0.04 & 0.36~$\pm$~0.03 & 1.05~$\pm$~0.09 & 2.86~$\pm$~0.20 & 0.41~$\pm$~0.03 \\
  044a\_6 & 26.66~$\pm$~1.71 & 4.78~$\pm$~0.53 & 0.52~$\pm$~0.06 & 0.34~$\pm$~0.03 & 1.00~$\pm$~0.09 & 2.86~$\pm$~0.22 & 0.44~$\pm$~0.04 \\
  044a\_9 & 8.69~$\pm$~1.00 & 5.02~$\pm$~1.01 & 0.56~$\pm$~0.11 & 0.29~$\pm$~0.05 & 0.86~$\pm$~0.14 & 2.86~$\pm$~0.39 & 0.51~$\pm$~0.07 \\
 044a\_12 & 4.38~$\pm$~0.28 & 4.24~$\pm$~0.76 & 0.74~$\pm$~0.12 & 0.28~$\pm$~0.06 & 0.73~$\pm$~0.07 & 2.74~$\pm$~0.19 & 0.50~$\pm$~0.06 \\
 044a\_15 & 3.48~$\pm$~0.29 & 2.98~$\pm$~0.68 & 0.58~$\pm$~0.17 & 0.22~$\pm$~0.07 & 0.61~$\pm$~0.08 & 2.21~$\pm$~0.20 & 0.44~$\pm$~0.07 \\
 044a\_wh & 295.18~$\pm$~5.79 & 2.95~$\pm$~0.10 & 0.43~$\pm$~0.01 & 0.61~$\pm$~0.02 & 1.79~$\pm$~0.05 & 2.86~$\pm$~0.07 & 0.35~$\pm$~0.01 \\
\nodata & & & & & & &
\enddata
\tablecomments{\scriptsize The fluxes in this table are  in units of $10^{-15}~\mathrm{ergs~s^{-1}~cm^{-2}}$ and have been corrected for internal dust attenuation. This table is available in its entirety in the online journal. A portion is shown here for guidance regarding its form and content.}
\label{Tab_Flux_0}
\end{deluxetable*}

\begin{figure*}[!ht]
\centering
\vspace*{-12mm}
\hspace*{-48mm}
\includegraphics[width=3.3\columnwidth]{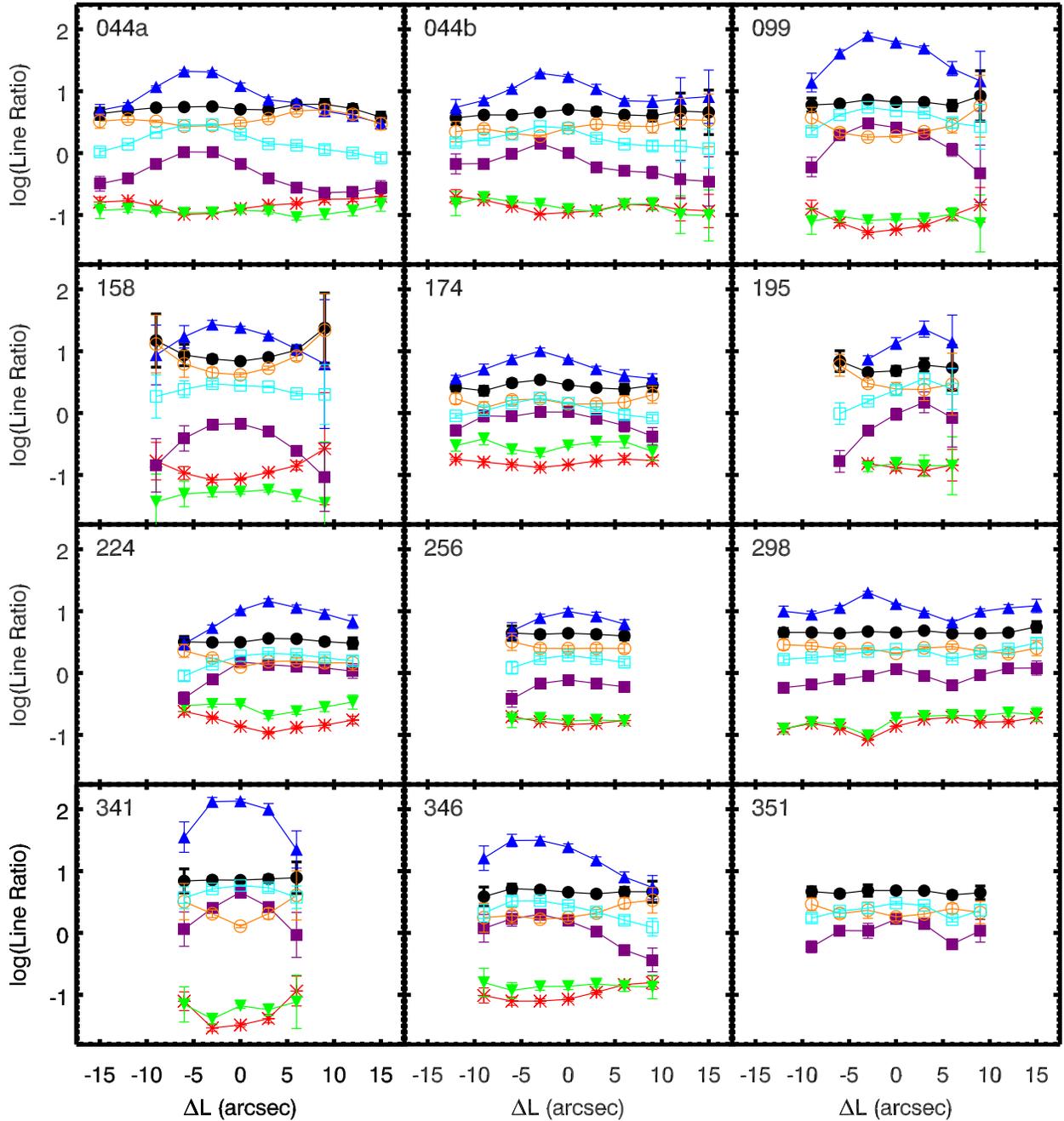}
\vspace*{-8mm}
\caption{Radial profiles of the spectral indices, $R23$ (black filled circles), $N2$ (red asterisks), $O3N2$ (blue filled triangles), $N2O2$ (green filled upside-down triangles), $R2$ (orange open circles), $R3$ (cyan open squares), and $O3O2$ (purple filled squares) in the individual \HII~regions. In each panel, the abscissa is the spatial coordinate in units of arcseconds, where $\Delta$L = 0 represents the center of one \HII~region defined in Section \ref{Sec_data}.} \label{Fig_R32}
\end{figure*}

In order to eliminate inevitable uncertainties propagated from the calibrations to the estimates during the derivation of the oxygen abundances and to analyze underlying causes of the diversity between the radial distributions in depth, we additionally plot radial profiles of the strong-line indices, $R23$, $N2O2$, $N2$, and $O3N2$, as well as $R2$ ( $\equiv$ $\log(\OII\lambda3727/\mathrm{H}\beta)$), $R3$ ( $\equiv$ $\log(\OIII\lambda\lambda4959,5007/\mathrm{H}\beta)$), and $O3O2$ ( $\equiv$ $\log(\OIII\lambda\lambda4959,5007/\OII\lambda3727)$, the other indices in the $R23$ diagnostics) for the \HII~regions in Figure \ref{Fig_R32}. Among the seven indices, $R2$, $R3$, and $N2$ represent relative fluxes for the collisionally excited lines $\OII\lambda3727$, $\OIII\lambda\lambda4959,5007$, and $\NII\lambda6583$, respectively, which consolidate the indices $R23$, $N2O2$, $O3N2$, and $O3O2$ by means of certain combinations.

As can be seen from this figure, $R3$, $O3N2$, and $O3O2$ uniformly decrease with the nebular radius, while $R2$ and $N2$ increase from the center to the edge. In contrast to such obvious gradients, $R23$ and $N2O2$, in most cases of our study, appear with shallow distributions. It is easy to understand that, when combining $R2$ and $R3$ into $R23$, the opposite radial variations in the two indices will compensate the offsets between each other along the nebular radius and consequently flatten the $R23$ profile; similarly, a combination of $N2$ and $R2$ in a ratio form will also have a counteracting effect on the radial variation and generate $N2O2$ with a gentle or even no slope in the profile. Nevertheless, if the collisionally excited line in $R2$ or $R3$ or $N2$ is strong enough to effectively dominate the consolidated index $R23$ or $N2O2$, the radial profile of $R23$ or $N2O2$ is still likely to exhibit a gradient. In our work, this situation occurs in the Nos. 158 and 174 \HII~regions, where $\OII\lambda3727$ is relatively overwhelming compared with other emission lines; $R23$ and $N2O2$ hence perform visible gradients as a result of the covariation with $R2$. On the other hand, the ratio of $R3$ to $N2$ (or $R2$) will inevitably produce an amplified radial variation, contributing the steepest gradients of $O3N2$ and $O3O2$ in all cases in Figure \ref{Fig_R32}. The radial distributions of $R23$, $N2O2$, $N2$,  $O3N2$, and $O3O2$ in our work coincide with earlier studies of the two \HII~regions NGC~595 and NGC~588 inside M33, observed with integral field spectrographs, where $O3N2$, $N2$, and $O3O2$ obviously correlate with nebular radii, whereas $R23$ and $N2O2$ are constantly distributed from centers to edges \citep[][]{2010MNRAS.402.1635R, 2011MNRAS.413.2242M}.

Figure \ref{Fig_R32} provides a straightforward interpretation of the radial variations in the oxygen abundances in Figure \ref{Fig_abun}, with a decrease or increase in relative fluxes for the collisionally excited lines at larger radii.\footnote{In case of a suspicion attributing the variations to lower signal-to-noise ratios at larger radii, we need to clarify that all the analyses in this work are based on relative fluxes (i.e., flux ratios, instead of absolute fluxes) on which an influence of noise ought to be random fluctuation rather than the systematic bias. Due to this fact, the decreasing or increasing trends with radii in the figures are not relevant to weaker signals when approaching the faint edge of a nebula.} Apart from the invalid data points from the Nos. 158 and 341 \HII~regions as marked in gray in Figure \ref{Fig_abun}, a complicated case appears in the index $R23$ from which the two estimates $R23_\mathrm{K99}$ and $R23_\mathrm{P05}$ clearly differ from each other in the radial fluctuation even if the offsets are neglected. In consideration of the consistency between $R23_\mathrm{K99}$, $N2O2_\mathrm{K02}$, and $N2O2_\mathrm{B07}$, the discrepancy of $R23_\mathrm{P05}$ from the three indices in the radial variations implies a possible problem lying in the $R23_\mathrm{P05}$ diagnostic. By comparing Figures \ref{Fig_abun} and \ref{Fig_R32} on a panel-by-panel basis, we find that the peaks or the valleys of $R23_\mathrm{P05}$ and $O3O2$ coincidentally occur one by one. Due to this behavior, we suspect that the $R23_\mathrm{P05}$ prescription is likely to overweigh $O3O2$ in terms of the parameterization and the formulation in Equations (\ref{Eq_R23l_P05}) and (\ref{Eq_R23h_P05}), compared to those in Equations (\ref{Eq_R23l_K99}) and (\ref{Eq_R23h_K99}) for $R23_\mathrm{K99}$ where the parameters $x$ and $y$ are actually different. We will attempt to make considerate inspections to validate or challenge this suspicion in our future work.

Besides the radial distributions presented above, the "strong-line" abundances for the \HII~regions with integrated measurements are also listed in Table \ref{Tab_Abun_int}. The deviation in the "integrated" abundances for the same \HII~region appears to be constant with uncertainties taken into account and not evidently related with other parameters. Integrated measurements of a large sample of \HII~regions will be conducted and systematic differences between the strong-line diagnostics will be investigated in a separate study (Lin et al. 2018, in preparation). In this current work, we compare the results of the measurements by adopting large and small apertures. Table \ref{Tab_Abun_cen} lists the "strong-line" abundances for the centers of the \HII~regions (i.e., by employing the 3 arcsec length aperture). A comparison between Tables \ref{Tab_Abun_int} and \ref{Tab_Abun_cen} manifests a trivial change within the range of error between the large- and small-aperture measurements, which suggests that the measurements with a small aperture are able to comply with the integrated measurements as long as the brightest part of the \HII~region is enclosed.

\section{\textbf{DISCUSSION}}\label{Sec_disc}

\begin{deluxetable*}{lrrrrrr}
\tabletypesize{\normalsize}
\tablecaption{Empirical Oxygen Abundances ($12 + \log(\mathrm{O/H})$) for the \HII~Regions with Integrated Measurements}
\tablewidth{0pc}
\tablehead{
\colhead{ID.} & \colhead{$R23_\mathrm{K99}$} & \colhead{$N2O2_\mathrm{K02}$} & \colhead{$R23_\mathrm{P05}$} & \colhead{$N2O2_\mathrm{B07}$} & \colhead{$O3N2_\mathrm{P04}$} & \colhead{$N2_\mathrm{P04}$}
}
\startdata
 044a & 8.655~$\pm$~0.013 & 8.635~$\pm$~0.012 & 8.317~$\pm$~0.016 & 8.177~$\pm$~0.010 & 8.356~$\pm$~0.005 & 8.321~$\pm$~0.005 \\
 044b & 8.726~$\pm$~0.017 & 8.702~$\pm$~0.017 & 8.426~$\pm$~0.020 & 8.236~$\pm$~0.015 & 8.359~$\pm$~0.007 & 8.315~$\pm$~0.007 \\
  099 & 8.597~$\pm$~0.018 & 8.513~$\pm$~0.025 & 8.384~$\pm$~0.020 & 8.084~$\pm$~0.018 & 8.164~$\pm$~0.008 & 8.187~$\pm$~0.006 \\
  158 & 8.419~$\pm$~0.025 & 8.205~$\pm$~0.049 & 7.971~$\pm$~0.032 & 7.915~$\pm$~0.021 & 8.303~$\pm$~0.005 & 8.269~$\pm$~0.006 \\
  174 & 8.882~$\pm$~0.013 & 8.879~$\pm$~0.014 & 8.583~$\pm$~0.017 & 8.408~$\pm$~0.014 & 8.458~$\pm$~0.008 & 8.372~$\pm$~0.009 \\
  195 & 8.663~$\pm$~0.071 & 8.677~$\pm$~0.064 & 8.330~$\pm$~0.089 & 8.214~$\pm$~0.055 & 8.370~$\pm$~0.026 & 8.345~$\pm$~0.027 \\
  224 & 8.849~$\pm$~0.011 & 8.885~$\pm$~0.011 & 8.589~$\pm$~0.014 & 8.414~$\pm$~0.012 & 8.421~$\pm$~0.007 & 8.373~$\pm$~0.007 \\
  256 & 8.753~$\pm$~0.027 & 8.770~$\pm$~0.022 & 8.402~$\pm$~0.034 & 8.298~$\pm$~0.021 & 8.433~$\pm$~0.012 & 8.384~$\pm$~0.013 \\
  298 & 8.732~$\pm$~0.008 & 8.781~$\pm$~0.006 & 8.431~$\pm$~0.009 & 8.309~$\pm$~0.006 & 8.396~$\pm$~0.003 & 8.378~$\pm$~0.004 \\
  341 & 8.585~$\pm$~0.027 & 8.320~$\pm$~0.063 & 8.387~$\pm$~0.030 & 7.967~$\pm$~0.032 & 8.063~$\pm$~0.013 & 8.095~$\pm$~0.012 \\
  346 & 8.728~$\pm$~0.023 & 8.678~$\pm$~0.025 & 8.464~$\pm$~0.027 & 8.214~$\pm$~0.022 & 8.310~$\pm$~0.010 & 8.273~$\pm$~0.009 \\
  351 & 8.726~$\pm$~0.018 & ... & 8.456~$\pm$~0.023 & ... & ... & ...
\enddata
\tablecomments{\small The table-head presents the six diagnostics adopted for estimating the oxygen abundances listed below.}
\label{Tab_Abun_int}
\end{deluxetable*}

\begin{deluxetable*}{lrrrrrr}
\tabletypesize{\normalsize}
\tablecaption{Empirical Oxygen Abundances ($12 + \log(\mathrm{O/H})$) for the Centers of the \HII~Regions}
\tablewidth{0pc}
\tablehead{
\colhead{ID.} & \colhead{$R23_\mathrm{K99}$} & \colhead{$N2O2_\mathrm{K02}$} & \colhead{$R23_\mathrm{P05}$} & \colhead{$N2O2_\mathrm{B07}$} & \colhead{$O3N2_\mathrm{P04}$} & \colhead{$N2_\mathrm{P04}$}
}
\startdata
 044a & 8.672~$\pm$~0.040 & 8.640~$\pm$~0.036 & 8.299~$\pm$~0.051 & 8.181~$\pm$~0.030 & 8.382~$\pm$~0.015 & 8.331~$\pm$~0.015 \\
 044b & 8.688~$\pm$~0.031 & 8.654~$\pm$~0.031 & 8.387~$\pm$~0.038 & 8.193~$\pm$~0.026 & 8.336~$\pm$~0.013 & 8.300~$\pm$~0.012 \\
  099 & 8.604~$\pm$~0.008 & 8.509~$\pm$~0.023 & 8.397~$\pm$~0.009 & 8.082~$\pm$~0.016 & 8.158~$\pm$~0.006 & 8.182~$\pm$~0.006 \\
  158 & 8.509~$\pm$~0.030 & 8.235~$\pm$~0.067 & 8.120~$\pm$~0.039 & 7.927~$\pm$~0.030 & 8.288~$\pm$~0.007 & 8.250~$\pm$~0.008 \\
  174 & 8.893~$\pm$~0.008 & 8.893~$\pm$~0.012 & 8.613~$\pm$~0.011 & 8.423~$\pm$~0.012 & 8.451~$\pm$~0.004 & 8.367~$\pm$~0.006 \\
  195 & 8.710~$\pm$~0.073 & 8.722~$\pm$~0.068 & 8.405~$\pm$~0.090 & 8.253~$\pm$~0.062 & 8.370~$\pm$~0.030 & 8.340~$\pm$~0.031 \\
  224 & 8.871~$\pm$~0.007 & 8.904~$\pm$~0.011 & 8.633~$\pm$~0.009 & 8.434~$\pm$~0.011 & 8.405~$\pm$~0.004 & 8.351~$\pm$~0.005 \\
  256 & 8.740~$\pm$~0.037 & 8.752~$\pm$~0.030 & 8.402~$\pm$~0.046 & 8.281~$\pm$~0.029 & 8.412~$\pm$~0.016 & 8.368~$\pm$~0.016 \\
  298 & 8.745~$\pm$~0.004 & 8.780~$\pm$~0.005 & 8.471~$\pm$~0.004 & 8.308~$\pm$~0.004 & 8.373~$\pm$~0.002 & 8.352~$\pm$~0.003 \\
  341 & 8.607~$\pm$~0.009 & 8.382~$\pm$~0.049 & 8.414~$\pm$~0.010 & 8.001~$\pm$~0.028 & 8.048~$\pm$~0.010 & 8.086~$\pm$~0.012 \\
  346 & 8.749~$\pm$~0.032 & 8.688~$\pm$~0.037 & 8.510~$\pm$~0.037 & 8.223~$\pm$~0.033 & 8.285~$\pm$~0.016 & 8.249~$\pm$~0.013 \\
  351 & 8.726~$\pm$~0.033 & ... & 8.493~$\pm$~0.039 & ... & ... & ...
\enddata
\tablecomments{\small The table-head presents the six diagnostics adopted for estimating the oxygen abundances listed below.}
\label{Tab_Abun_cen}
\end{deluxetable*}

In this work, we confirm the systematic offsets between the empirical oxygen abundances separately estimated with the observation- and model-based prescriptions. Physical origins of the offsets are still unclear. Aside from inappropriate treatments of ionization structures in theoretical models, biased sampling of emission-line sources in the observation-based calibrations is suspected to be a possible cause \citep{2003ApJ...591..801K, 2007ApJ...656..186B, 2008ApJ...681.1183K}. The observational relationships between the strong lines and the $T_\mathrm{e}$-sensitive auroral lines are often obtained with integrated measurements of \HII~regions or star-forming galaxies. However, due to temperature fluctuations within an ionized nebula, detected auroral lines in most cases actually indicate higher temperature and thus lower metallicity than measured strong lines for the same data point in calibration diagrams. Therefore, with the peak $T_\mathrm{e}$ referencing the averaged strong lines in a nebula, the empirical diagnostics are supposed to underestimate true oxygen abundances at certain degrees. The solution of this problem is to recalibrate the empirical abundance indicators by measuring both auroral and strong lines for identical positions inside one \HII~region.

The radial variations in the empirical oxygen abundances, or more directly, the strong-line indices, are significant fruits of our investigation. It is not realistic for actual oxygen abundances to vary so sharply on a nebular scale ($\lesssim 330$ pc). Therefore, our results imply the existence of additional parameters affecting the widely used abundance indicators. The sensitivities of $\OII\lambda3727$, $\OIII\lambda\lambda4959,5007$, and $\NII\lambda6583$ to ionization levels suggest the ionization parameter as a candidate to underlie the spectral indices, in addition to the oxygen abundance. Among these collisionally excited lines, $\OIII\lambda\lambda4959,5007$ are more efficiently excited in regions with a high degree of ionization, while $\OII\lambda3727$ and $\NII\lambda6583$ are more related with a low-ionization state \citep{1996ias..book.....E, 2006agna.book.....O}. In this situation, a decrease in the ionization parameter is supposed to raise $R2$ and $N2$, to diminish $R3$, $O3N2$, and $O3O2$, and to ineffectively change $R23$ and $N2O2$, if intrinsic metallicities are constant. At the same time, ionization states are postulated to inherently decline at larger radii of an ionized nebula. In support of this postulation, 3-D nebular models configured with such an ionization structure have successfully reproduced 2-D multi-wavelength features of individual \HII~regions \citep{2011MNRAS.412..675P, 2014A&A...566A..12P}. As a consequence, the radial gradients of $N2$ and $O3N2$ in Figure \ref{Fig_R32} are consistent with the natural distribution of the ionization parameter within one \HII~region and considered to be an imprint of the ionization parameter rather than the the oxygen abundance, whereas the flat profiles of $R23$ and $N2O2$ manifest the robustness of the two indices against ionization variations, which has been disclosed by several other observations of \HII~regions in the Milky Way and the Magellanic Clouds \citep{2000ApJ...537..589K, 2000ApJ...539..687O} as well as local galaxies \citep{2007ApJ...656..186B, 2016ApJ...816...40J}, also in agreement with expectations of photoionization models \citep{2002ApJS..142...35K, 2013ApJS..208...10D}. Likewise, the different sensitivities of the emission lines to the radially decreasing ionization parameter offer an interpretation of the disparity between the gradients displayed in each panel of Figure \ref{Fig_lgL0}, i.e., steeper for $\OIII\lambda\lambda4959,5007$ and shallower for $\OII\lambda3727$ and $\NII\lambda6583$.

The reliability of $R23$ and $N2O2$, at least when the ionization parameter varies, which is the predominant situation in our study, is also reflected in our work by consistent fluctuations in the radial profiles of model-based $R23_\mathrm{K99}$ and $N2O2_\mathrm{K02}$ shown in each panel of Figure \ref{Fig_abun}, albeit the same emission line $\OII\lambda3727$ is located in the numerator of one index ($R23$) but the denominator of the other ($N2O2$). In contrast to the proximate overlap between $R23_\mathrm{K99}$ and $N2O2_\mathrm{K02}$, $R23_\mathrm{P05}$ deviates from $N2O2_\mathrm{B07}$ by even up to 0.4 dex, as depicted in Figure \ref{Fig_abun}. As a commonly accepted probe of ionization levels, $O3O2$ is combined with $R23$ into the diagnostics for correcting the influences of the ionization parameter on $R23$. However, in our work, the flat distributions of $R23$ in Figure \ref{Fig_R32} and the intensive fluctuations of $R23_\mathrm{P05}$ in Figure \ref{Fig_R32} suggest that, $R23$, neutralized by $R2$ and $R3$ though incompletely in many cases, appears not so sensitive to the degree of ionization as $N2$ and $O3N2$, and the ionization correction in the $R23_\mathrm{P05}$ diagnostic needs to be reexamined in a high variety of ionization states.

The failure of applying $N2$ and $O3N2$ to various ionization states has been predicted by photoionization models \citep[e.g.,][]{2002ApJS..142...35K}, yet there has not been direct observational evidence demonstrating the correlation of $N2$ or $O3N2$ with the ionization parameter. Nevertheless, observations of \HII~regions in the Magellanic Clouds have revealed more spatially extended contours for low-ionization emission lines than high-ionization ones \citep{2012ApJ...755...40P}, which actually implies potential deviation in abundance estimates from ionization-sensitive emission lines such as $\OIII\lambda\lambda4959,5007$ and $\NII\lambda6583$. Despite this drawback of $N2$ and $O3N2$, in the case that ionization states are not quite variable, $N2$ and $O3N2$ are still applicable, in particular to heavily dust-obscured regions since both of the indices are independent of dust attenuation.

In a short summary, through spatially resolved spectrophotometry of 11 \HII~regions in NGC~2403 and thereby naturally sampling various ionization states, our study corroborates the theoretical expectation and offers the observational evidence of the similarities and differences between the empirical diagnostics of the oxygen abundance. Further confirmation of the interpretations of the discrepancies, such as the temperature fluctuations and the ionization diversities, will rest on comparison of the empirical estimates with the $T_\mathrm{e}$-based oxygen abundance, which requires successful detection of the auroral lines. However, in this present work, for NGC~2403 with the metallicity $> 0.5 Z_\odot$ \citep{2013ApJ...775..128B}, the auroral lines are undetectable with the 2.16 m telescope, which hampers us to recalibrate the strong-line indices or obtain the internal distributions of $T_\mathrm{e}$ for the \HII~regions. In future work, we plan to target metal-poor \HII~regions in nearby galaxies or in the Milky Way, where the auroral lines are observable, aimed at an in-depth exploration on the factors affecting the empirical abundance diagnostics.

\begin{figure*}[!ht]
\centering
\hspace*{-10mm}
\includegraphics[width=1.5\columnwidth]{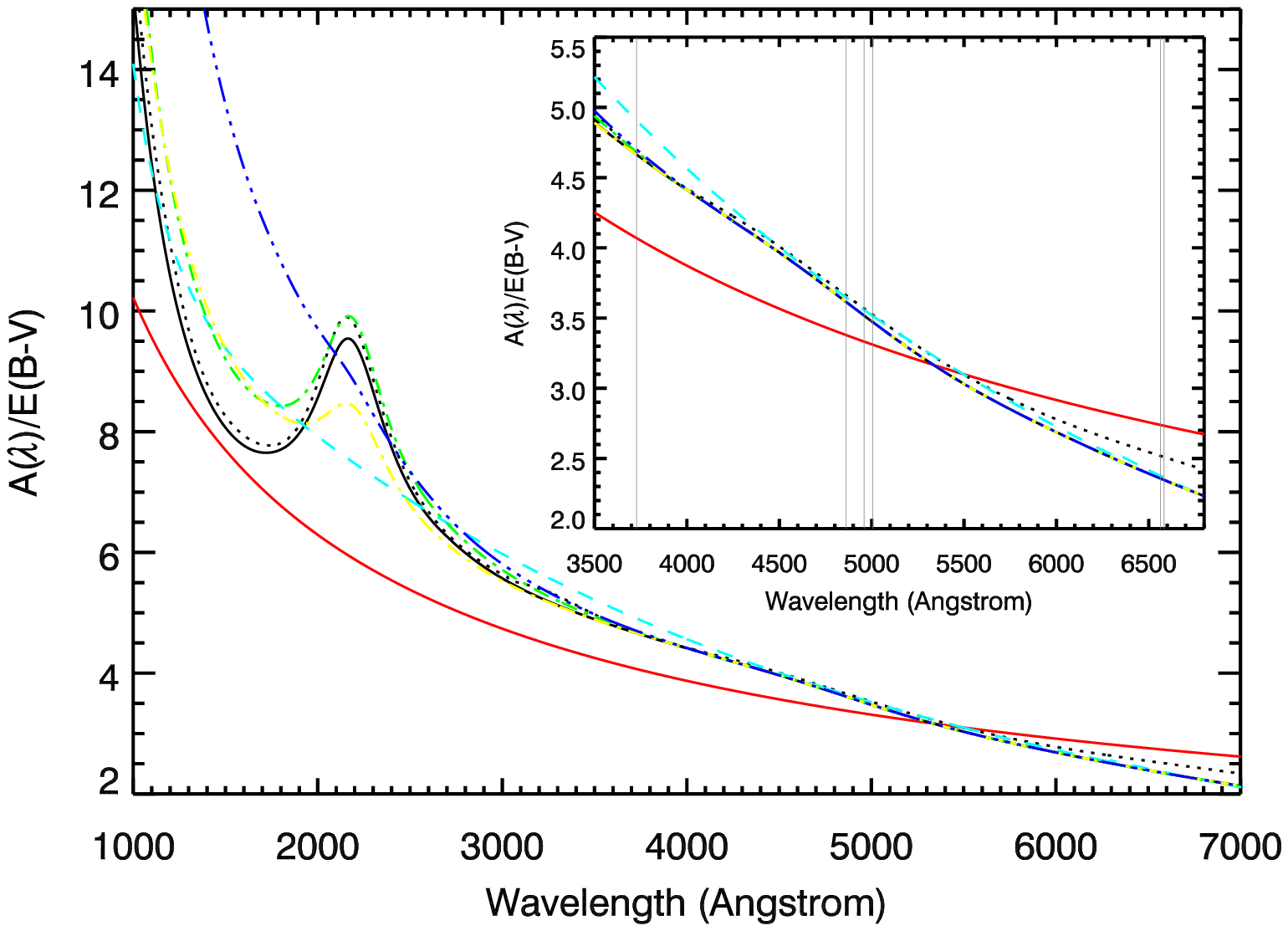}
\caption{Seven attenuation/extinction curves in a wavelength range from UV to optical bands with $R_\mathrm{V} = 3.1$, including the \citet{1999PASP..111...63F} curve (black solid line), the \citet{2000ApJ...533..682C} curve (cyan dashed line), the \citet{1989ApJ...345..245C} curve (black dotted line), the \citet{2003ApJ...594..279G} curve for the Large Magellanic Cloud (green dotted-dashed line), the \citet{2003ApJ...594..279G} curve for the supershell area in the Large Magellanic Cloud (yellow dotted-dashed line), the \citet{2003ApJ...594..279G} curve for the Small Magellanic Cloud (blue triple-dotted-dashed line), and the \citet{2000ApJ...539..718C} curve of $\lambda^{-0.7}$ (red solid line). The internal panel is plotted to highlight the difference between these curves in the optical range relevant to our work, where the gray solid lines mark the wavelength-positions for the emission lines, $\OII\lambda3727$, H$\beta$, $\OIII\lambda\lambda4959,5007$, H$\alpha$, and $\NII\lambda6583$, from left to right on the horizontal axis.} \label{Fig_AttCurv}
\end{figure*}


\acknowledgments

We are grateful to Robert C. Kennicutt, Jr., for providing substantive comments on the manuscript. We appreciate the careful review and the instructive comments offered by the anonymous referee, which have improved the paper significantly. This work is supported by the National Natural Science Foundation of China (NSFC, Nos. 1320101002, 11421303, 11433005, 11590782, 11603007, 11703063, and U1731104) and the National Key R\&D Program of China (2015CB857004 and 2017YFA0402600). We acknowledge the support of the staff of the XingLong 2.16 m telescope. This work has been partially supported by the Open Project Program of the Key Laboratory of Optical Astronomy, National Astronomical Observatories, Chinese Academy of Sciences. We thank Zhou Fan and Wei Zhang for their hospitality during the periods of the observations at National Astronomical Observatories of China. Ye-Wei Mao thanks Monica Relano-Pastor for helpful discussion at the early stage of this work. Ye-Wei Mao acknowledges the support of the Start-up Fund of GuangZhou University. This research has made use of NASA's Astrophysics Data System.

\appendix

\section{\textbf{AN INSPECTION OF DIFFERENT ATTENUATION CURVES}}\label{App}

\begin{figure*}[!ht]
\centering
\vspace*{-12mm}
\hspace*{-48mm}
\includegraphics[width=3.3\columnwidth]{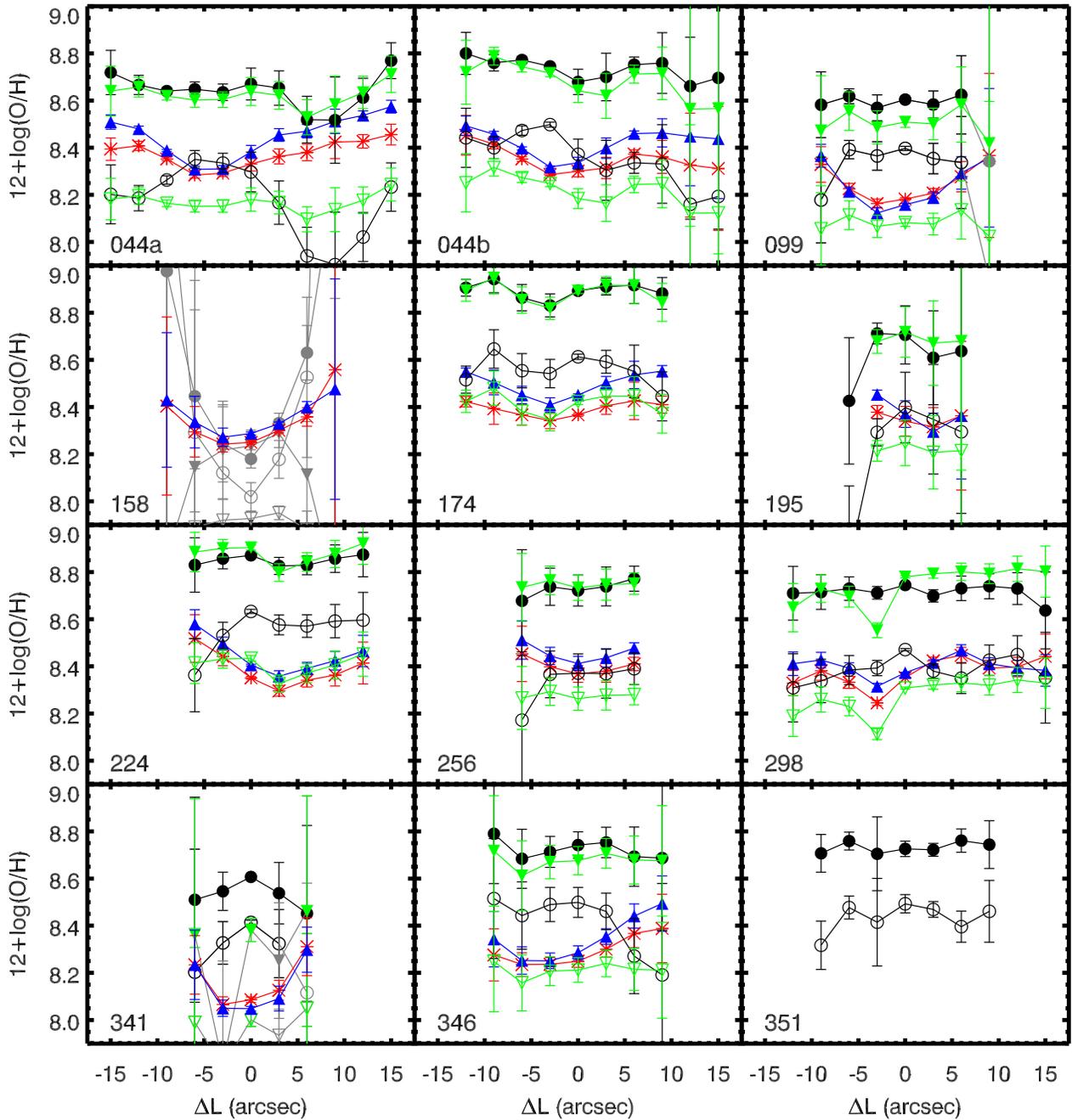}
\vspace*{-8mm}
\caption{Same diagrams with Figure \ref{Fig_abun} but the fluxes for the emission lines are corrected for the internal dust attenuation by adopting the \citet{2000ApJ...539..718C} curve of $\lambda^{-0.7}$.} \label{Fig_abun_dis_cf}
\end{figure*}

Attenuation curves, depicting dust attenuation as a function of wavelength, are a necessary material for compensating dust attenuation and recovering intrinsic spectra in observational astrophysics. At present, attenuation curves have been found to be various in form. In this work, we apply the \citet{1999PASP..111...63F} attenuation curve to the \HII~regions observed in NGC~2403. The basic frame of this curve consists of three components parameterized with five coefficients \citep{1988ApJ...328..734F, 1990ApJS...72..163F}. This three-component parameterization has been tested to be feasible by a number of studies of the Milky Way \citep[e.g.,][]{2004ApJ...602..291W, 2005ApJ...625..167S, 2009ApJ...705.1320G}, the Magellanic Clouds \citep[e.g.,][]{2003ApJ...594..279G, 2005ApJ...630..355C, 2012A&A...541A..54M}, and galaxies from low to high redshifts \citep[e.g.,][]{1994A&A...291....1R, 2012ApJ...753...82Z, 2015ApJ...815...14C}. The \citet{1999PASP..111...63F} curve is the specific parameterization by mean values for a number of sightlines in the Milky May, as is adopted in our work.

Another form of attenuation curves is provided by \citet{1989ApJ...345..245C}, parameterized with $R_\mathrm{V}$ only, and widely used for Galactic extinction correction. We employed the \citet{1989ApJ...345..245C} curve for correcting Galactic extinction in this work. The \citet{1989ApJ...345..245C} curve is in agreement with the \citet{1999PASP..111...63F} curve when $R_\mathrm{V} \sim 3.1$. However, the $R_\mathrm{V}$-dependent property for the attenuation curve has never been discovered in galaxies other than the Milky Way. Thus, application of the \citet{1989ApJ...345..245C} curve to extraGalactic environments is likely to render a mistake.

Through an investigation of galaxies as a whole, \citet{2000ApJ...533..682C} have obtained an attenuation curve expressed by a polynomial equation and exhibiting a smooth shape; a similarly featureless attenuation curve in a power-law form of $\lambda^{-0.7}$ has been produced with modeling by \citet{2000ApJ...539..718C}. Both of the \citet{2000ApJ...533..682C} and \citet{2000ApJ...539..718C} curves are interpreted to be a statistical approximation of "age-selective attenuation" which describes heavier dust obscuration for younger stellar populations \citep{2000ApJ...542..710G, 2007MNRAS.375..640P}. Consequently, they are more suitable for statistic censuses of galaxies with integrated measurements. For studies of certain objects, these two curves appear to oversimplify the actual properties of dust obscuration.

In most cases, people are not able to ascertain an attenuation curve in a straightforward way but only presume one instead. The applicabilities of the attenuation curves presented above will help to make the choice in specific studies. In order to better illustrate respective characteristics of various attenuation curves, we perform a visual comparison by displaying seven typical attenuation curves, including the \citet{1999PASP..111...63F} curve adopted in our work, the \citet{1989ApJ...345..245C} curve, the \citet{2000ApJ...533..682C} curve, the \citet{2000ApJ...539..718C} curve, and the curves for the Magellanic Clouds developed by \citet{2003ApJ...594..279G}, in Figure \ref{Fig_AttCurv}. As can be obviously seen, the most striking distinction of these attenuation curves lies in the wavelength range shorter than 2500 \AA, which is due to differences in not only the slope of the curves but also the strength of the 2175 \AA~bump \citep[see][for elaborated influences of altering attenuation curves on UV-band observations]{2014ApJ...789...76M}. However, in the optical range, where our study is carried out, there is a high degree of consistency between all the curves except the \citet{2000ApJ...539..718C} one. In order to further inspect the influence of replacing the \citet{1999PASP..111...63F} curve with the \citet{2000ApJ...539..718C} curve on the results, we replot the radial profiles of the oxygen abundances on the basis of the correction for internal dust attenuation with the \citet{2000ApJ...539..718C} curve in Figure \ref{Fig_abun_dis_cf}, which shows that the change in the attenuation curve leads to very slight shifts for a few data points and does not affect our conclusions.
\\

\end{document}